\documentclass[twocolumn,preprintnumbers,amsmath,amssymb, superscriptaddress]{revtex4-1}
\usepackage[utf8]{inputenc}
\usepackage[english]{babel}
\usepackage{graphicx}
\usepackage{mathtools}
\usepackage{color}

\newcommand*{\mat}[1]{\boldsymbol{#1}}
\newcommand*{\coord}[1]{\mathbf{#1}}
\newcommand*{\vecr}{\coord{r}}
\newcommand*{\vecx}{\coord{x}}
\newcommand*{\vecy}{\coord{y}}

\usepackage{slashed}
\newcommand{\pino}{\ensuremath\mathord{\mspace{2mu}\declareslashed{}{\scriptstyle\diagup}{0}{0.1}{\pi}\!\slashed{\pi}}}

\newcommand*{\crea}[1]{\hat{#1}^{\dagger}}
\newcommand*{\anni}[1]{\hat{#1}^{\protect\vphantom{\dagger}}}

\DeclareMathOperator{\Fourier}{\mathcal{F}}
\DeclareMathOperator{\Imag}{\mathfrak{Im}}
\DeclareMathOperator{\Real}{\mathfrak{Re}}
\DeclareMathOperator{\Trace}{\mathrm{Tr}}

\newcommand*{\binteg}[3]{\int^{\mathrlap{#3}}_{\mathrlap{#2}}\ud{#1}\,}
\newcommand*{\integ}[1]{\int\!\!\!\:\ud{#1}\:}
\newcommand*{\iinteg}[2]{\integ{#1}\!\!\!\integ{#2}}

\newcommand*{\braket}[2]{\langle{#1}|{#2}\rangle}
\newcommand*{\bigbraket}[2]{\bigl\langle{#1}\big|{#2}\bigr\rangle}
\newcommand*{\brakket}[3]{\langle{#1}|{#2}|{#3}\rangle}

\newcommand*{\du}{\partial}
\newcommand*{\e}{\mathrm{e}}
\newcommand*{\eqspace}{\phantom{{} = {}}}
\newcommand*{\half}{\frac{1}{2}}
\newcommand*{\im}{\mathrm{i}}
\newcommand*{\isDefinedAs}{\coloneqq}
\newcommand*{\ud}{\mathrm{d}}
\newcommand*{\Wcal}{\mathcal{W}}

\usepackage{hyperref}

\allowdisplaybreaks[4]

\begin{document}

\title{Response calculations based on an independent particle system with the exact one-particle density matrix: polarizabilities}
\author{K.J.H. Giesbertz}
\email{k.j.h.giesbertz@vu.nl}
\affiliation{Section Theoretical Chemistry, VU University, De Boelelaan 1083, 1081 HV Amsterdam, The Netherlands}

\author{O.V. Gritsenko}
\affiliation{Section Theoretical Chemistry, VU University, De Boelelaan 1083, 1081 HV Amsterdam, The Netherlands}
\affiliation{Department of Chemistry, Pohang University of Science and Technology, San 31, Hyojadong, Namgu, Pohang 790-784, South-Korea}

\author{E.J. Baerends}
\affiliation{Section Theoretical Chemistry, VU University, De Boelelaan 1083, 1081 HV Amsterdam, The Netherlands}
\affiliation{Department of Chemistry, Pohang University of Science and Technology, San 31, Hyojadong, Namgu, Pohang 790-784, South-Korea}
\affiliation{Department of Chemistry, Faculty of Science, King Abdulaziz University, Jeddah 21589, Saudi Arabia}

\date{\today}

\begin{abstract}
Recently, we have demonstrated that the problems finding a suitable adiabatic approximation in time-dependent one-body reduced density matrix functional theory can be remedied by introducing an additional degree of freedom to describe the system: the phase of the natural orbitals [Phys.\ Rev.\ Lett.~\textbf{105}, 013002 (2010), J.~Chem.\ Phys.~\textbf{133}, 174119 (2010)]. In this article we will show in detail how the frequency-dependent response equations give the proper static limit ($\omega\to0$), including the perturbation in the chemical potential, which is required in static response theory to ensure the correct number of particles. Additionally we show results for the polarizability for H$_2$ and compare the performance of two different two-electron functionals: the phase-including Löwdin--Shull functional and the density matrix form of the Löwdin--Shull functional.
\end{abstract}

\maketitle

\section{Introduction}
Time-dependent one-body reduced density matrix functional theory (TD1MFT), provides an interesting alternative to time-dependent density functional theory (TDDFT). The description of excitations while breaking bonds goes catastrophically wrong in TDDFT~\cite{GiesbertzBaerendsGritsenko2008, GiesbertzBaerends2008}. Double excitations are absent in adiabatic TDDFT, which therefore fails for excited state potential energy surfaces (PES), which rapidly acquire double excitation character at elongated bond lengths, as demonstrated for the lowest excited $\Sigma_g^+$ surface of H$_2$~\cite{GiesbertzBaerendsGritsenko2008} and for the lowest $\Pi_u$ surface (b$^1\Pi_u$) in N$_2$~\cite{NeugebauerBaerends2004}. Both these types of excitations are feasible with TD1MFT~\cite{GiesbertzBaerendsGritsenko2008, GiesbertzPernalGritsenko2009}, and also charge-transfer excitations can be described without difficulty with TD1MFT.  For practical calculations, the use of an adiabatic approximation is mandatory. However, in the case of TD1MFT the standard adiabatic (SA) approximation as is usually employed in TDDFT, leads to some unphysical results: the occupation numbers become time-independent as demonstrated in Refs~\cite{PernalGiesbertzGritsenko2007, Appel2007, GiesbertzPernalGritsenko2009, GiesbertzGritsenkoBaerends2010a, AppelGross2010} and the frequency-dependent linear response equations~\cite{GiesbertzGritsenkoBaerends2010a, GiesbertzGritsenkoBaerends2010b} are in the static limit ($\omega\to0$) not equal to the static response equations~\cite{CioslowskiPernal2001, PernalBaerends2006}.

To avoid these problems, an alternative adiabatic approximation was proposed~\cite{PernalGiesbertzGritsenko2007, PernalCioslowski2007}. This adiabatic approximation assumes that the occupation numbers instantaneously relax, so are determined by their ground state equations. This approximation has been given the more descriptive name instantaneous occupation number relaxation (IONR) approximation by Requist and Pankratov~\cite{RequistPankratov2010}. Unfortunately, the occupation numbers are still not dynamic variables, which impairs the description of some dynamic phenomena. An example are the ``diagonal'' double excitations, like the $(1\sigma_g)^2 \to (1\sigma_u)^2$ excitation of the first  $^1\Sigma_g^+$ excited state in H$_2$ at elongated bond lengths ($\gtrsim 5$ Bohr)~\cite{GiesbertzPernalGritsenko2009}.

An alternative way to resolve these issues due to the adiabatic approximation in TD1MFT is to extend the description of the system with additional variables, which can be introduced and treated as the natural orbital (NO) phase factors. The phase factors of the NOs are not defined, since they are the eigenfunctions of the 1RDM. To distinguish these combinations of NOs and phase factors from the NOs, they are named phase including NOs (PINOs). It has been demonstrated that the explicit treatment of the phase of the PINOs leads to a number of significant improvements over TD1MFT in the SA approximation~\cite{GiesbertzGritsenkoBaerends2010a, GiesbertzGritsenkoBaerends2010b, GiesbertzGritsenkoBaerends2012b, MeerGritsenkoGiesbertz2013}. Not only a correct static limit is recovered as in the IONR approximation, but also the occupation numbers become truly dynamic, so also diagonal double excitations can be described such as the lowest $^1\Sigma_g^+$ excitation of elongated H$_2$, and off-diagonal double excitations as in the lowest state of N$_2$. Furthermore, an explicit treatment of the phase factors implies that also the functionals will be dependent on the phases of the PINOs. This has the advantage that the energy expression of the Löwdin--Shull wavefunction for two-electron systems~\cite{LowdinShull1956} can be written as a phase including NO functional (called PILS). Treating these additional phase variables in the response formalism of TD1MFT leads to a large improvement in performance over the phase-independent form, which is obtained by casting the Löwdin--Shull energy in the form of a  density matrix functional (called DMLS).

In this article we will focus on the calculation of dynamic polarizabilities. In particular, the static limit ($\omega\to0$) will be important. Therefore, after an introduction to the theory, we will show in detail how the static response equations are recovered. As a demonstration, we will show results of polarizability calculations for the simple two-electron system H$_2$. Also we study the effects of removing the majority of the virtual-virtual pairs from the polarizability calculation to decrease the computational cost. (With virtual we mean a weakly occupied PINO.) We obtain similar encouraging results as for the excitation energies and oscillator strengths~\cite{GiesbertzGritsenkoBaerends2012b, MeerGritsenkoGiesbertz2013}.

\section{PINO functional theory}
In TD1MFT, not only the density is used to describe the system of interest, but the complete one-body reduced density matrix (1RDM), which can be defined for a state $\Psi$ as
\begin{align*}
\gamma(\vecx,\vecx';t) \isDefinedAs \brakket{\Psi}{\crea{\psi}_{\text{H}}(\vecx't)\anni{\psi}_{\text{H}}(\vecx t)}{\Psi},
\end{align*}
where $\crea{\psi}_{\text{H}}(\vecx t)$ and $\anni{\psi}_{\text{H}}(\vecx t)$ are the usual field operators in the Heisenberg picture and $\vecx = \vecr\sigma$ is a combined space-spin coordinate.
The central idea of TD1MFT is that all the quantities of interest can be defined as a functional of the 1RDM, in particular the action functional
\begin{align*}
A[\gamma] \isDefinedAs \binteg{t}{0}{T} \brakket{\Psi[\gamma](t)}{\im\du_t - \hat{H}(t)}{\Psi[\gamma](t)}.
\end{align*}
In particular for the case of local potentials this statement holds, since by the Runge--Gross theorem~\cite{RungeGross1984} and its extensions~\cite{Leeuwen1999, Leeuwen2001, RuggenthalerLeeuwen2011}, the action is a functional of the density, which is trivially recovered from the 1RDM as its diagonal in coordinate representation, $\gamma(\vecx,\vecx;t)$. For full variation over the 1RDM, an extension of the Runge--Gross theorem to 1RDMs and non-local potentials is required.

The 1RDM is hermitian, so it can be diagonalized
\begin{align*}
\gamma(\vecx,\vecx';t) = \sum_kn_k(t)\phi_k(\vecx t)\phi_k^*(\vecx't).
\end{align*}
The eigenvalues, $n_k(t)$, are called the (natural) occupation numbers and the eigenfunctions, $\phi_k(\vecx t)$, are called the natural orbitals (NOs)~\cite{Lowdin1955a}. The occupation numbers and NOs are equivalent to the 1RDM, so functionals can be defined in terms of the occupation numbers and NOs in stead of the 1RDM, $A[\gamma] = A[\{\phi,n\}]$. However, since the NOs are eigenfunctions of a hermitian operator, the phase of the NOs is not defined and therefore, the functionals $A[\{\phi,n\}]$, are not allowed to depend on them. Unfortunately, this phase independence leads to problems in the SA approximation as has been demonstrated in Refs~\cite{RequistPankratov2010, GiesbertzGritsenkoBaerends2010a, GiesbertzGritsenkoBaerends2010b, RequistPankratov2011} and also extends to higher order reduced density matrices~\cite{Requist2012}. Therefore, it has been proposed to go beyond TD1MFT and to use this phase information as well. The action is assumed to be a functional of the PINOs and occupation numbers, $A[\{\pino,n\}]$. To be able to derive some useful equations of motions (EOMs) from the action, it is split in an non-interacting part, $A_0$ based on a non-interacting ensemble, and a remainder, $A_{\text{Hxc}}$, as
\begin{align*}
A[\{\pino,n\}] = A_0[\{\pino,n\}] + A_{\text{Hxc}}[\{\pino,n\}],
\end{align*}
where the non-interacting part is defined as
\begin{align*}
A_0 \isDefinedAs \binteg{t}{0}{T} \sum_rn_r(t)\brakket{\pino_{\!r}(t)}{\im\du_t - \hat{h}(t)}{\pino_{\!r}(t)}.
\end{align*}
The one-body hamiltonian, $\hat{h}(t)$, is the one-body part of the interacting system and contains the usual kinetic and potential terms.

Useful expressions for the EOMs of the PINOs and occupation numbers can now be derived by taking functional derivatives. However, one has to keep in mind the boundary terms of the upper time-limit as pointed out by Vignale~\cite{Vignale2008}. Therefore, we introduce
\begin{align*}
\delta\Wcal[\{\pino,n\}] &\isDefinedAs \im\braket{\Psi(T)}{\delta\Psi(T)} - \delta A_{\text{Hxc}} \\*
&\hphantom{{}\isDefinedAs{}} {} -
\im\sum_rn_r(T)\braket{\pino_{\!r}(T)}{\delta\pino_{\!r}(T)},
\end{align*}
which allows us to formulate stationarity of the action as
\begin{align*}
\delta A_0 - \delta\Wcal = \im\sum_rn_r(T)\braket{\pino_{\!r}(T)}{\delta\pino_{\!r}(T)}.
\end{align*}
Taking functional derivatives with respect to the occupation numbers gives the EOM for the PINO phases
\begin{align}\label{eq:EOMpinoPhase}
\im\bigbraket{\pino_{\!k}(t)}{\dot{\pino}_{\!k}(t)}
&= h_{kk}(t) + \frac{\delta\Wcal}{\delta n_k(t)}.
\end{align}
Taking functional derivatives with respect to the PINOs and taking ortho-normality into account one recovers the EOM of the 1RDM
\begin{multline}\label{eq:EOM1RDM}
\im\Bigl[\dot{n}_k(t)\delta_{kl} + 
\bigl(n_l(t) - n_k(t)\bigr)\bigbraket{\pino_{\!k}(t)}{\dot{\pino}_{\!l}(t)}\Bigr] \\
= \bigl(n_l(t) - n_k(t)\bigr)h_{kl}(t) + 
\bigl(\Wcal^{\dagger}_{kl}(t) - \Wcal^{\vphantom{\dagger}}_{kl}(t)\bigr),
\end{multline}
where we defined
\begin{align*}
\Wcal_{kl}[\{\pino,n\}](t) \isDefinedAs \integ{\vecx}\frac{\delta \Wcal}{\delta\pino_{\!k}(\vecx t)}\pino_{\!l}(\vecx t).
\end{align*}
The EOMs can be recombined, to give an EOM for the occupation numbers
\begin{align*}
\im\dot{n}_k(t) = \Wcal^{\dagger}_{kk}(t) - \Wcal^{\vphantom{\dagger}}_{kk}(t)
\end{align*}
and an EOM for the PINOs
\begin{align*}
\im\du_t\pino_{\!k}(\vecx t) = \bigl(\hat{h}(t) + \hat{v}^{\text{PINO}}(t)\bigr)\pino_{\!k}(\vecx t),
\end{align*}
where the effective PINO potential is defined by its matrix elements as
\begin{align*}
v^{\text{PINO}}_{kl}(t) \isDefinedAs \begin{cases}
\vphantom{\Bigg[}
\dfrac{\Wcal^{\dagger}_{kl}(t) - \Wcal_{kl}(t)}{n_l(t) - n_k(t)}	&\text{for $k \neq l$} \\
\vphantom{\Bigg[}
\dfrac{\delta\Wcal}{\delta n_k(t)}						&\text{for $k = l$}.
\end{cases}
\end{align*}
The standard adiabatic (SA) approximation can now simply be defined as $\Wcal[\{\pino,n\}] \approx W[\{\pino,n\}] $, where $W[\{\pino,n\}]$ is the ground state functional for the two-body part of the energy. In practice, this will be one of our approximations to the exact ground state functional.

To formulate the linear response equations, it is convenient to expand the time-dependent perturbation in the PINOs in the stationary PINOs as~\cite{GiesbertzGritsenkoBaerends2012b}
\begin{align}\label{eq:dUdef}
\delta\pino_{\!k}(\vecx t) = \sum_r\pino_{\!r}(\vecx) e^{-\im\epsilon_r t}\delta\mat{U}_{rk}(t).
\end{align}
To preserve the orthonormality of the $\{\pino_{\!k}(\vecx t)\}$, the matrix $\delta \mat{U}$ has to be antihermitian, $\delta \mat{U} =-\delta \mat{U}^\dagger$. 
The off-diagonal elements in $\delta\mat{U}(t)$ have the following simple relation to the perturbation in the 1RDM
\begin{align*}
\delta\gamma_{kl}(t) = \delta_{kl}\delta n_k(t) + (n_l - n_k)\delta U_{kl}(t).
\end{align*}
We can therefore use the off-diagonal $\delta U_{kl}$ instead of the $\delta \gamma_{kl}$, which can be numerically more convenient when the occupation numbers are very small or very close to each other. Note that the diagonal $\delta U_{kk}$ are purely imaginary and (being the components of $\delta \pino_{\!k}(\vecx t)$ along the stationary $\pino_{\!k}(\vecx)e^{-\im\epsilon_k t}$) describe the phase of the time-dependent PINOs.  Further, it is convenient to introduce the following notation to indicate the real and imaginary parts of the matrices in the time-domain
\begin{align*}
f^{R/I}_{kl}(t) &\isDefinedAs [\Real/\Imag] f_{kl}(t).
\end{align*}
We define the vector of diagonal $\delta \mat{U}$ elements as $\delta U^D_k \isDefinedAs \delta U_{kk}^I$. Assuming that the reference PINOs, $\pino_{\!k}(\vecx)$, are real we can write the adiabatic linear response equations as~\cite{PhD-Giesbertz2010, GiesbertzGritsenkoBaerends2010b}
\begin{subequations}\label{eq:time-dep_resp}
\begin{align}
\delta\dot{\gamma}^R_{kl}(t) &= \sum_{a>b}\bigl(A_{kl,ba} + A_{kl,ab}\bigr)
\frac{\delta\gamma^I_{ab}(t)}{n_b-n_a} \notag \\*
&\; {} +
\sum_a A_{kl,aa}\delta U^D_a(t) + (n_l-n_k)\delta v^I_{kl}(t), \\
\delta\dot{n}_k(t) &= \sum_{a>b}\bigl(A_{kk,ba} + A_{kk,ab}\bigr)
\frac{\delta\gamma^I_{ab}(t)}{n_b-n_a} \notag \\*
&\eqspace {} +
\sum_aA_{kk,aa}\delta U^D_a(t), \\
-\delta\dot{\gamma}^I_{kl}(t) &= \sum_{a>b}\bigl(A_{kl,ba} - A_{kl,ab}\bigr)
\frac{\delta\gamma^R_{ab}(t)}{n_b-n_a} \notag \\*
&\eqspace {} +
\sum_aC_{kl,a}\delta n_a(t) + (n_l-n_k)\delta v^R_{kl}(t), \\
-\delta\dot{U}^D_k(t) &= 2\sum_{a>b}C_{ab,k}\frac{\delta\gamma^R_{ab}(t)}{n_b-n_a} \notag \\*
&\eqspace {} +
2\sum_a\overline{W}_{k,a}\delta n_a(t) + \delta v_{kk}(t),
\end{align}
\end{subequations}
where we introduced the following response matrices
\begin{subequations}
\begin{align}\label{eq:Adef}
A_{kl,ba} &\isDefinedAs (n_b-n_a)(h_{ka}\delta_{bl} - \delta_{ka}h_{bl}) \notag \\*
&\eqspace {} +
\integ{\vecx}\biggl(\frac{\du\bigl(W^{\dagger}_{kl} - W^{\phantom{\dagger}}_{kl}\bigr)}{\du \pino_b(\vecx)}\pino_a(\vecx) \\*
&\eqspace\hphantom{\integ{\vecx}\biggl(} {} - 
\frac{\du\bigl(W^{\dagger}_{kl} - W^{\phantom{\dagger}}_{kl}\bigr)}{\du \pino^*_a(\vecx)}\pino^*_b(\vecx)\biggr), \notag \\
\label{eq:Cdef}
C_{kl,a} &\isDefinedAs h_{kl}(\delta_{al} - \delta_{ka}) + 
\frac{\du\bigl(W^{\dagger}_{kl} - W^{\phantom{\dagger}}_{kl}\bigr)}{\du n_a}, \\
\label{eq:WbarDef}
\overline{W}_{k,a} &\isDefinedAs \half\frac{\du^2 W}{\du n_k\du n_a}.
\end{align}
\end{subequations}
The frequency-dependent response equations can be obtained by simply taking the Fourier transform. The resulting set of linear equations can be represented in the following matrix form~\cite{PhD-Giesbertz2010, GiesbertzGritsenkoBaerends2010b}
\begin{widetext}
\begin{align}\label{eq:fullRespMatrix}
\mat{\chi}^{-1}_{\text{SA}}(\omega)
\begin{pmatrix}
\delta\mat{\gamma}^R(\omega)	\\
\delta\mat{n}(\omega)		\\
\im\delta\mat{U}^I(\omega)	\\
\im\delta\mat{U}^D(\omega)/2
\end{pmatrix}
\isDefinedAs
\begin{pmatrix}
\omega\mat{1}_M				&\mat{0} 				&-\mat{A}_{MM}^+		&-\mat{A}_{Mm}^+	\\
\mat{0}						&\omega\mat{1}_m		&-\mat{A}_{mM}^+		&-\mat{A}_{mm}^+	\\
-\mat{N}^{-1}\mat{A}^-\mat{N}^{-1}	&-\mat{N}^{-1}\mat{C}	&\omega\mat{1}_M		&\mat{0}			\\
-\mat{C}^T\mat{N}^{-1}			&-\mat{\overline{W}}		& \mat{0}				&\omega\mat{1}_m
\end{pmatrix}
\begin{pmatrix}
\delta\mat{\gamma}^R(\omega)	\\
\delta\mat{n}(\omega)		\\
\im\delta\mat{U}^I(\omega)	\\
\im\delta\mat{U}^D(\omega)/2
\end{pmatrix}
=
\begin{pmatrix}
\mat{N}\im\delta\mat{v}^I(\omega)	\\
\mat{0}						\\
\delta\mat{v}^R(\omega)			\\
\delta\mat{v}^D(\omega)/2
\end{pmatrix},
\end{align}
\end{widetext}
where we used the following definitions
\begin{align*}
N_{kl,ba} &\isDefinedAs (n_l - n_k)\delta_{ka}\delta_{bl}, \\
A^{\pm}_{kl,ba} &\isDefinedAs A_{kl,ba} \pm A_{kl,ab}.
\end{align*}
The inverse response matrix on the left-hand side of Eq.~\eqref{eq:fullRespMatrix} is blocked as $(M,m,M,m)$, with $M \isDefinedAs m(m-1)/2$ and $m$ is the size of the basis set. The subblocks of $\mat{A}^+$ which are diagonal in either the column index ($A^+_{kl,aa}$) or the row index ($A^+_{kk,ab}$) or both ($A^+_{kk,aa}$) are indicated with appropriate subscripts, $A^+_{Mm}$, $A^+_{mM}$ or $A^+_{mm}$, respectively.

From these linear response equations (in the SA approximation) it is straightforward to find the excitation energies by solving the homogenous equation, or to find one-body-one-body response functions such as the polarizability which is defined as minus the real part of the dipole-dipole response function. Before we turn our attention to the results for the polarizability of H$_2$, we will first show in detail how these response equations behave in the $\omega\to0$ limit and how the static response equations~\cite{PernalBaerends2006} are fully recovered, contrary to what happens in the SA approximation in TD1MFT~\cite{PernalGiesbertzGritsenko2007, GiesbertzGritsenkoBaerends2010b}.

\section{Static limit}
It has already been observed~\cite{GiesbertzGritsenkoBaerends2010b} that the static limit ($\omega\to0$) has to be treated very carefully, lest discrepancies arise between the static response equations~\cite{PernalBaerends2006} (which should be recovered at exactly $\omega =0$) and the small but finite $\omega$ values occurring when the limit $\omega \to 0$ is taken in frequency dependent calculations.  This issue is addressed here.  We will see that the perturbation by a spatially constant time-dependent potential plays a special role in the linear response equations, so before we can consider the static limit $\omega\to 0$, we need to investigate the response to such potentials first, i.e.\ to perturbations of the following type
\begin{align*}
\delta v_{kl}(t) = \delta v(t)\delta_{kl}.
\end{align*}
The constant time-dependent potential should play a similar role as in TDDFT where it only affects the time-dependent phase factor of the wavefunction and does not lead to a response of the density~\cite{RungeGross1984}. Therefore, we expect that the constant time-dependent potential only induces a global change of PINO phase factors and no response of the 1RDM. Hence
\begin{align*}
\qquad\qquad\qquad\delta\gamma_{kl}(t) &= 0 &&\forall_{k,l}, \qquad \\
\qquad\qquad\qquad-\delta\dot{U}^D_k(t) &= \delta v(t) &&\forall_k \qquad\qquad\qquad
\end{align*}
should be a solution of the time-dependent response equations~\eqref{eq:time-dep_resp}. Using the fact that $\delta U^D_k(0)=0$ independent of $k$, and therefore all $\delta U^D_k(t)$ equal at all times according to the second equation above, this solution is readily verified, if the following sum-rule is satisfied by the response matrix $\mat{A}$
\begin{align}\label{eq:Asumrule}
\sum_aA_{kl,aa} = 0 \qquad \forall_{k,l}.
\end{align}
The derivation of this sum-rule is quite technical and has been deferred to the Appendix~\ref{ap:AsumRule}. Taking the Fourier transform, we find the frequency dependent counterpart of the equations above  
\begin{align*}
\delta\gamma_{kl}(\omega) &= 0 \quad\forall_{k,l}, \\*
\omega\,\im\delta U^D_k(\omega) &= \Fourier\bigl[-\delta\dot{U}^D_k\bigr](\omega) = \delta v(\omega)
\quad \forall_k.
\end{align*}
This solution implies that in the static limit, $\omega \to 0$, $\delta U^D_k(\omega)$ diverges if $\delta v(0) \ne 0$, such that
\begin{align*}
\lim_{\omega\to0}\omega\,\im\delta U^D_{k}(\omega) = \delta v(0) \quad \forall_k.
\end{align*}
For general perturbing potentials, divergence can be avoided by explicitly subtracting a time\slash{}frequency-dependent constant $\delta \epsilon(\omega)$ from the perturbing potential, so $\delta \mat{v}^D(\omega) \to \delta \mat{v}^D(\omega) - \delta\epsilon(\omega)\mat{1}_m$.  As argued before, it is allowed to add a constant shift to the potential, since it does not lead to any physical response of the system.

The constant part of the potential for $\omega = 0$ can be obtained by requiring $\im\delta\mat{U}^D(\omega \to 0)$ not to diverge, so the $\omega\,\im\delta\mat{U}^D(\omega)$ term will now vanish from the last set of equations of the frequency dependent response~\eqref{eq:fullRespMatrix} for $\omega \to 0$. By multiplying from the left by the occupation numbers $\mat{n}^T$, we find that the constant part of the potential, $\delta\epsilon(0)$, has to satisfy
\begin{align*}
0 = \sum_kn_k\bigl(\delta v^D_{k}(0) - \delta\epsilon(0)\bigr),
\end{align*}
where we used that the response matrices $\mat{C}$ and $\overline{\mat{W}}$ satisfy the following sum-rules (see Appendix~\ref{ap:CandWsumRules})
\begin{align}\label{eq:CandWbarSumRules}
\sum_a C_{kl,a}n_a &= 0 &
&\text{and}&
\sum_a \overline{W}_{k,a}n_a &= 0.
\end{align}
The simplest definition of the constant part of the potential at finite frequencies consistent with the zero frequency result is
\begin{align}\label{eq:deltaepsilon(omega)}
\delta\epsilon(\omega) = \frac{1}{N}\sum_k n_k\delta v_{kk}(\omega).
\end{align}
With the explicit elimination of the constant part of the potential, we can take the zero frequency limit of the frequency dependent response equations, without having to worry about a possible divergence of $\im\delta\mat{U}^D(\omega)$. In particular for real perturbations, $\im\delta\mat{v}^I = \mat{0}$, the response equations in the static limit reduce to
\begin{subequations}\label{eq:PINO_stat}
\begin{align}\label{eq:PINO_stat_dphi}
\mat{A}^-\delta\mat{U}^R(0) + \mat{C}\delta\mat{n}(0) + \mat{N}\delta\mat{v}^R(0) &= \mat{0}, \\
\label{eq:PINO_stat_dn}
2\,\mat{C}^T\delta\mat{U}^R(0) + 2\,\mat{\overline{W}}\delta\mat{n}(0) + \delta\mat{v}^D(0)
&= \delta \epsilon(0)\,\mat{1}_m.
\end{align}
\end{subequations}
These static response equations are identical to the static response equations derived earlier, cf.\ Eqns (40) and~(41) of Ref.~\cite{PernalBaerends2006} (see Eqns (21a) and~(21b) of Ref.~\cite{GiesbertzGritsenkoBaerends2010b} in the present notation). This immediately reveals that the $\delta \epsilon(0)$ constant we have introduced is in fact the first order change in the Lagrange multiplier introduced in Refs~\cite{PernalBaerends2006, GiesbertzGritsenkoBaerends2010b} to enforce a constant number of electrons. This fits in with our present introduction of $\delta \epsilon(0)$ because of the necessity of keeping the number of electrons constant by avoiding divergence of $\im\delta\mat{U}^D(\omega \to 0)$. Eq.~\eqref{eq:deltaepsilon(omega)} can be regarded as a frequency dependent generalization of the perturbation in the Lagrange multiplier to enforce the correct number of electrons as derived in Ref.~\cite{GiesbertzGritsenkoBaerends2010b}.
We note that the explicit elimination of the constant part of the potential is not only useful to demonstrate that we correctly recover the static response equations in the $\omega \to 0$ limit, but it is also useful in practical calculations. Without the elimination of the frequency-dependent constant, $\im\delta\mat{U}^D(\omega)$ would diverge for $\omega \to 0$ causing problems in numerical calculations. The $\delta\epsilon(\omega)$ prevents these complications.

Note that for a smooth $\omega\to0$ limit it is important that the functional truly depends on the PINO phases. If the functional $W$ does not depend on the PINO phase factors, $\alpha_k$, which true TD1MFT functionals do not do, then
\begin{align*}
0 = \frac{\ud W}{\ud \alpha_k} &= \integ{\vecx}\left(
\frac{\du W}{\du \pino_{\!k}(\vecx)}\frac{\ud\pino_{\!k}(\vecx)}{\ud\alpha_k} +
\frac{\du W}{\du \pino^*_{\!k}(\vecx)}\frac{\ud\pino^*_{\!k}(\vecx)}{\ud\alpha_k}\right) \\
&= \im\integ{\vecx}\left(\frac{\du W}{\du \pino_{\!k}(\vecx)}\pino_{\!k}(\vecx) - 
\frac{\du W}{\du \pino^*_{\!k}(\vecx)}\pino^*_{\!k}(\vecx)\right) \\
&= \im\bigl(W^{\vphantom{\dagger}}_{kk} - W^{\dagger}_{kk}\bigr).
\end{align*}
Since $\bigl(W^{\vphantom{\dagger}}_{kk} - W^{\dagger}_{kk}\bigr)$ vanishes identically, also all its derivatives vanish in the response matrix $\mat{A}$, especially we find that $\mat{A}^+_{mM} = \mat{0}$ and $\mat{A}^+_{mm} = \mat{0}$ and also $\mat{A}^+_{Mm} = \mat{0}$, since $\mat{A}$ is hermitian in the sense that $A^{\vphantom{*}}_{kl,ba} = A^*_{ab,lk}$ (Appendix~\ref{ap:AsumRule}). Hence, the coupling with the phase factors in the first three sets of PINO response equations~\eqref{eq:fullRespMatrix} is lost. Since also $\mat{A}^+_{mM}$ vanishes, we find that the second Eq.\  \eqref{eq:fullRespMatrix}reduces to $\omega\delta\mat{n}(\omega) = \mat{0}$, so for finite frequencies we have additionally that $\delta\mat{n}(\omega) = \mat{0}$, which underlines the problem of lack of occupation number response in true 1RDM functionals (phase independent) mentioned before. The second of the Eqns \eqref{eq:fullRespMatrix} can then be left out of the response equations. So for a phase independent functional $W$, the adiabatic PINO response equations become
\begin{align*}
\begin{pmatrix}
\omega\mat{1}_M				&-\mat{A}^+_{MM} \\
-\mat{N}^{-1}\mat{A}^-\mat{N}^{-1}	&\omega\mat{1}_M
\end{pmatrix}
\begin{pmatrix}
\delta\mat{\gamma}^R(\omega) \\
\im\delta\mat{U}^I(\omega)
\end{pmatrix}
=
\begin{pmatrix}
\mat{N}\im\delta\mat{v}^I(\omega) \\
\delta\mat{v}^R(\omega)
\end{pmatrix}.
\end{align*}
which means they have reverted back to the TD1MFT response in the SA approximation~\cite{GiesbertzBaerendsGritsenko2008, GiesbertzPernalGritsenko2009}. The perturbation in the PINO phase factors (which have an arbitrary initial value) can be solved afterwards from the last equation as
\begin{align*}
\omega\,\im\delta\mat{U}^D(\omega)
= 2\mat{C}^T\delta\mat{U}^R(\omega) + \delta\mat{v}^D(\omega) - \delta\epsilon(\omega)\mat{1}_m,
\end{align*}
though this not of any practical use, since in this case the PINO phases do not couple to any physical observables.

If there is no phase-dependence in the functional,  the standard adiabatic approximation leads to a special situation at zero frequency, $\omega = 0$: the conditions $\omega\delta\mat{n}(\omega) = \mat{0}$ do not imply anymore that there is no change in the occupation numbers. This causes a jump in the solutions (a discontinuity from $\omega \to 0$ and $\omega = 0$). At $\omega=0$ the perturbation in the $\delta \mat{n}$ becomes well defined, since the coupling to the PINO phase factors now also disappears completely from the last set of response equations, so $\delta\mat{\gamma}^R(0)$ and $\delta\mat{n}(0)$ are now determined from the static linear response equations~\eqref{eq:PINO_stat}. The perturbation in the imaginary part of the 1RDM is directly related to the imaginary part of the perturbing potential as
\begin{align*}
\im\delta\mat{\gamma}^I(0) = 
\mat{N}\im\delta\mat{U}^I(0) = -\mat{N}\bigl(\mat{A}^+_{MM}\bigr)^{-1}\mat{N}\im\delta\mat{v}^I(0).
\end{align*}
For a demonstration of this jump in the frequency dependent response when a phase invariant functional is used, we refer the reader to Ref.~\cite{GiesbertzGritsenkoBaerends2010b}, where the discontinuity has been illustrated with the $\alpha_{zz}(\omega)$ polarizability of the HeH\textsuperscript{+} system with the DMLS functional~\eqref{eq:DMLSdef}.

\section{Polarizability}
The polarizability is defined as the negative of the dipole-dipole response, so it can simply be obtained by evaluating the dipole response due to a dipolar field as perturbation. Using the response equations [Eq.~\eqref{eq:fullRespMatrix}], one can immediately obtain an expression for the polarizability. However, we assumed that the stationary PINOs are real, so only $\delta\mat{\gamma}^R(\omega)$ and $\delta\mat{n}(\omega)$ are required to evaluate the induced dipoles. Therefore, $\im\delta\mat{U}^I(\omega)$ and $\im\delta\mat{U}^D(\omega)$ are not of interest and can be eliminated from the equations. Shuffling the terms around in the equations, we obtain the following expression for the polarizability
\begin{align*}
\alpha_{\eta,\nu}(\omega) = 
-\begin{pmatrix}
2\mat{\nu}^T_M 	&\mat{\nu}^T_m
\end{pmatrix}
\left[\omega^2\mat{1} - \mat{A}^+\mat{D}\right]^{-1}\mat{A}^+
\begin{pmatrix}
\mat{\eta}_M \\
\mat{\eta}_m/2
\end{pmatrix},
\end{align*}
for $\eta, \nu = x,y,z$ and $\mat{\eta}_M$, $\mat{\eta}_m$ denote off-diagonal and diagonal matrix elements of $\eta_{kl} \isDefinedAs \brakket{\pino_k}{\eta}{\pino_l}$ respectively and similarly for $\mat{\nu}$.
Further, we introduced the following matrix
\begin{align*}
\mat{D} \isDefinedAs \begin{pmatrix}
\mat{N}^{-1}\mat{A}^-\mat{N}^{-1}	&\mat{N}^{-1}\mat{C} \\
\mat{C}^T\mat{N}^{-1}				&\mat{\overline{W}}
\end{pmatrix}.
\end{align*}
In principle the sum-rule of the matrix $\mat{A}$~\eqref{eq:Asumrule} should take care of the constant of the dipole matrix elements. Unfortunately, this sum-rule is usually not well satisfied in practice due to finite numerical precision, which leads to erratic behavior near $\omega = 0$. Therefore, we explicitly project out the constant shift $\delta\epsilon$ in practical calculations, so for the diagonal dipole elements we use
\begin{align*}
\nu_{kk} \to \nu_{kk} - \frac{1}{N}\sum_rn_r \nu_{rr}.
\end{align*}
In this article we will restrict ourselves to singlet two-electron systems. It has often been claimed that the exact 1MFT functional is known for such systems. However, there are two forms in use which are identical to each other for real (PI)NOs in the ground state, but lead to quite different results in the time-dependent regime. The one closest to the original Löwdin--Shull expression~\cite{LowdinShull1956} is the phase including Löwdin--Shull (PILS)
\begin{align*}
W^{\text{PILS}}[\{\pino,n\}] = \half\sum_{rs}\sqrt{n_rn_s}w_{rrss},
\end{align*}
where the two-electron integrals are defined as
\begin{align}
w_{klrs} \isDefinedAs \iinteg{\vecx}{\vecy}
\pino^*_{\!k}(\vecx)\pino^*_{\!l}(\vecy)w(\vecx,\vecy)\pino_{\!r}(\vecy)\pino_{\!s}(\vecx).
\end{align}
The integral $w_{rrss}=\braket{rr}{ss}$ is not a normal exchange integral $K_{rs}=\braket{rs}{sr}$ since the complex conjugation differs. With real functions there is of course no difference in numerical value. The PILS functional is not a proper 1MFT functional, since its expression is not phase invariant, i.e.\ it includes a dependence on the PINO phases through the two-electron integrals. By changing the phase of the PINOs the sign of the contribution to the sum can be influenced.

To derive a proper 1MFT functional, i.e.\ to make the functional phase-invariant, one can swap two indices in the integral which gives the following expression
\begin{align}\label{eq:DMLSdef}
W^{\text{DMLS}}[\{\phi,n\}] = \half\sum_{rs}f^*_rf_s\sqrt{n_rn_s}w_{rsrs}.
\end{align}
The integral $w_{rsrs} = \braket{rs}{sr} = K_{rs}$ is now the usual exchange integral, which does not depend on the phase of the orbitals anymore. Since the phase of the orbitals can not influence the signs of the contributions anymore, explicit phase factors $\{f_k = \pm 1\}$ need to be included in the phase-invariant form. This functional is now a proper 1MFT functional for fixed $\{f_k\}$ and is therefore named density matrix Löwdin--Shull (DMLS) functional.

\begin{figure}[t]
  \includegraphics[width=\columnwidth]{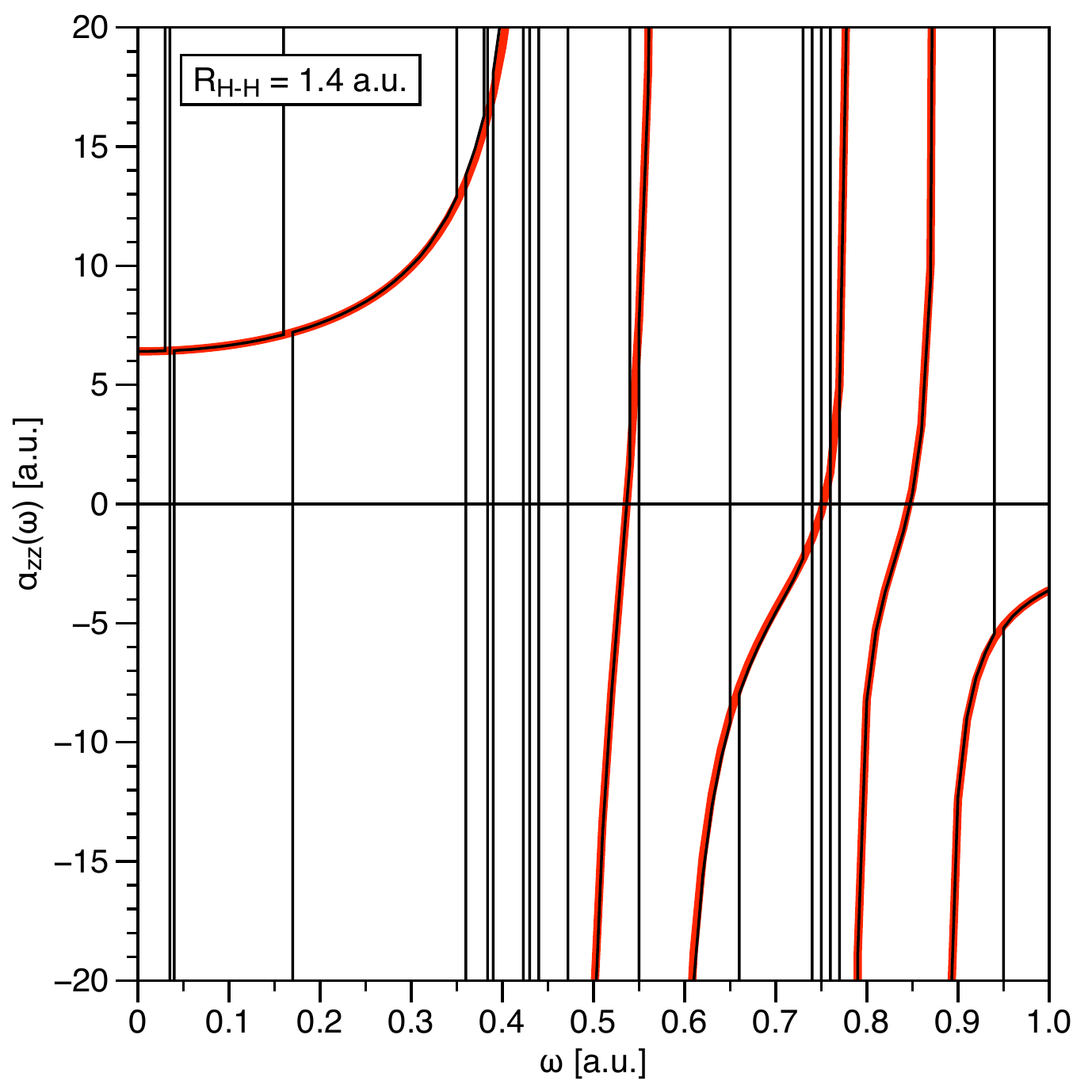}
  \includegraphics[width=\columnwidth]{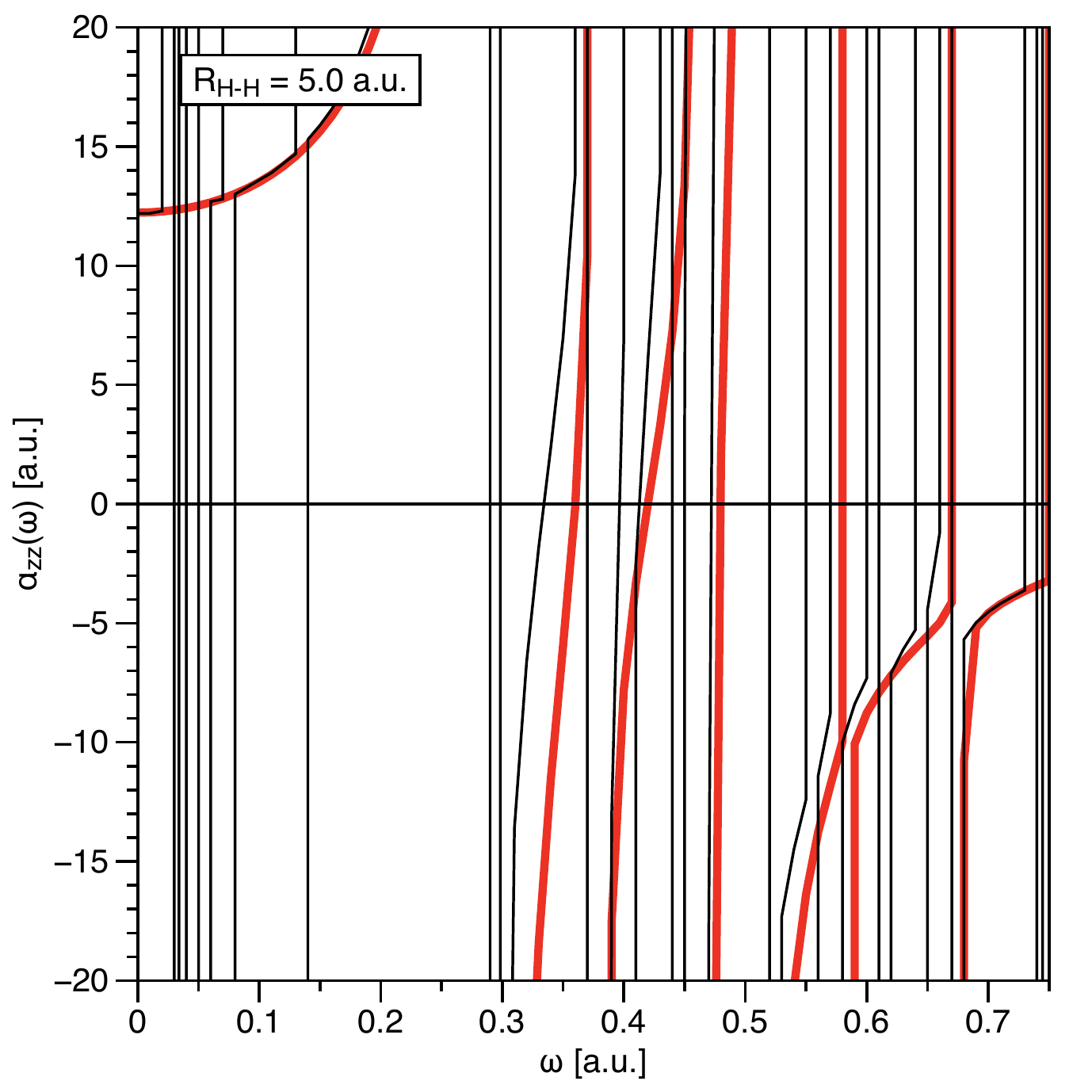}
  \caption{The $\alpha_{zz}(\omega)$ component of the polarizability tensor for H$_2$ at interatomic distances of 1.4 and 5.0 Bohr in an aug-cc-pVTZ basis. The exact (full CI) and PILS results exactly coincide. Exact (full CI) \& PILS: thick (red) lines; DMLS: thin (black) lines. All matrix elements have been taken into account for the PINO response calculation.}
  \label{fig:H2_zz_all}
\end{figure}

An additional advantage of using a normal exchange integral is that we have with DMLS an example of a so-called JK-only functional. JK-only functionals have been the first trial functionals. The derivation of Müller~\cite{Muller1984} starts from the Hartree-Fock exchange and therefore always uses just K integrals for the exchange-correlation part. The derivation of Buijse and Baerends~\cite{PhD-Buijse1991, BuijseBaerends2002} does not determine the phases and affords either K or L integrals. Until now the choice for K integrals (hence JK-only NO functionals) has almost universally been made since it seems to naturally connect to the correlation-less Hartree-Fock model and because they are pure 1RDM functionals. For ground state calculations (real orbitals) the choice of phases was immaterial anyway. However, in time-dependent 1RDM\slash{}PINO functional theory this subtle difference expresses itself in horrific results for dynamic properties in the case of the DMLS functional, in particular many low lying spurious excitation energies are produced~\cite{GiesbertzGritsenkoBaerends2010b}. These spurious excitations show up as very narrow divergencies (poles) in the polarizability, visible as ``spikes" in Fig.~\ref{fig:H2_zz_all} for the $\alpha_{zz}(\omega)$ component for H$_2$ at interatomic distances of 1.4 and 5.0 Bohr respectively. The calculation has been done in an aug-cc-pVTZ basis~\cite{cc-pVT/QZ_H_B-Ne_aug_H} and all response matrix elements $\delta \gamma_{kl}$ have been taken into account (see below  for tests with reduced numbers of $\delta \gamma_{kl}$ matrix elements). As a reference, also the full configurations interaction (CI) results are shown in red.

\begin{figure}[t]
  \includegraphics[width=\columnwidth]{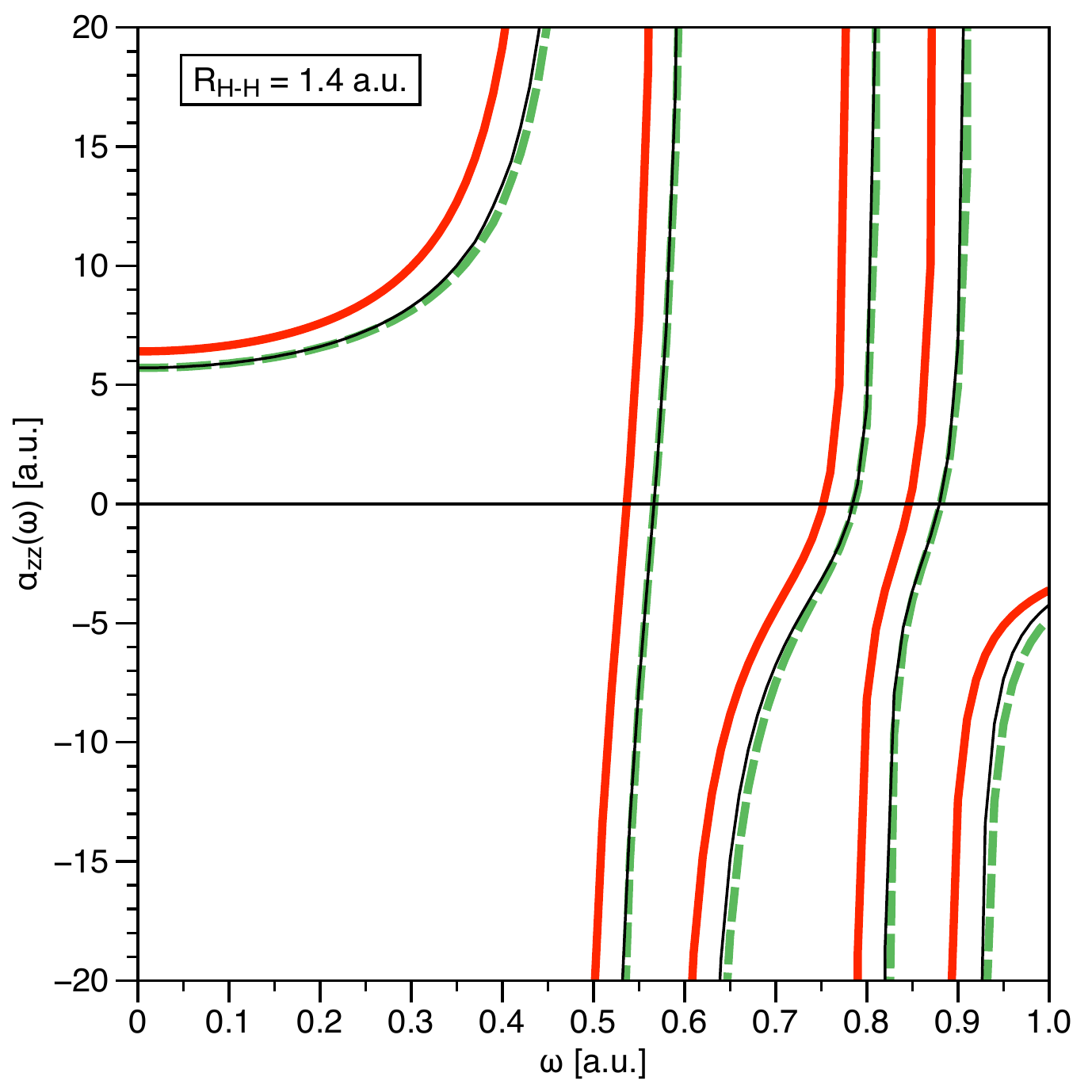}
  \includegraphics[width=\columnwidth]{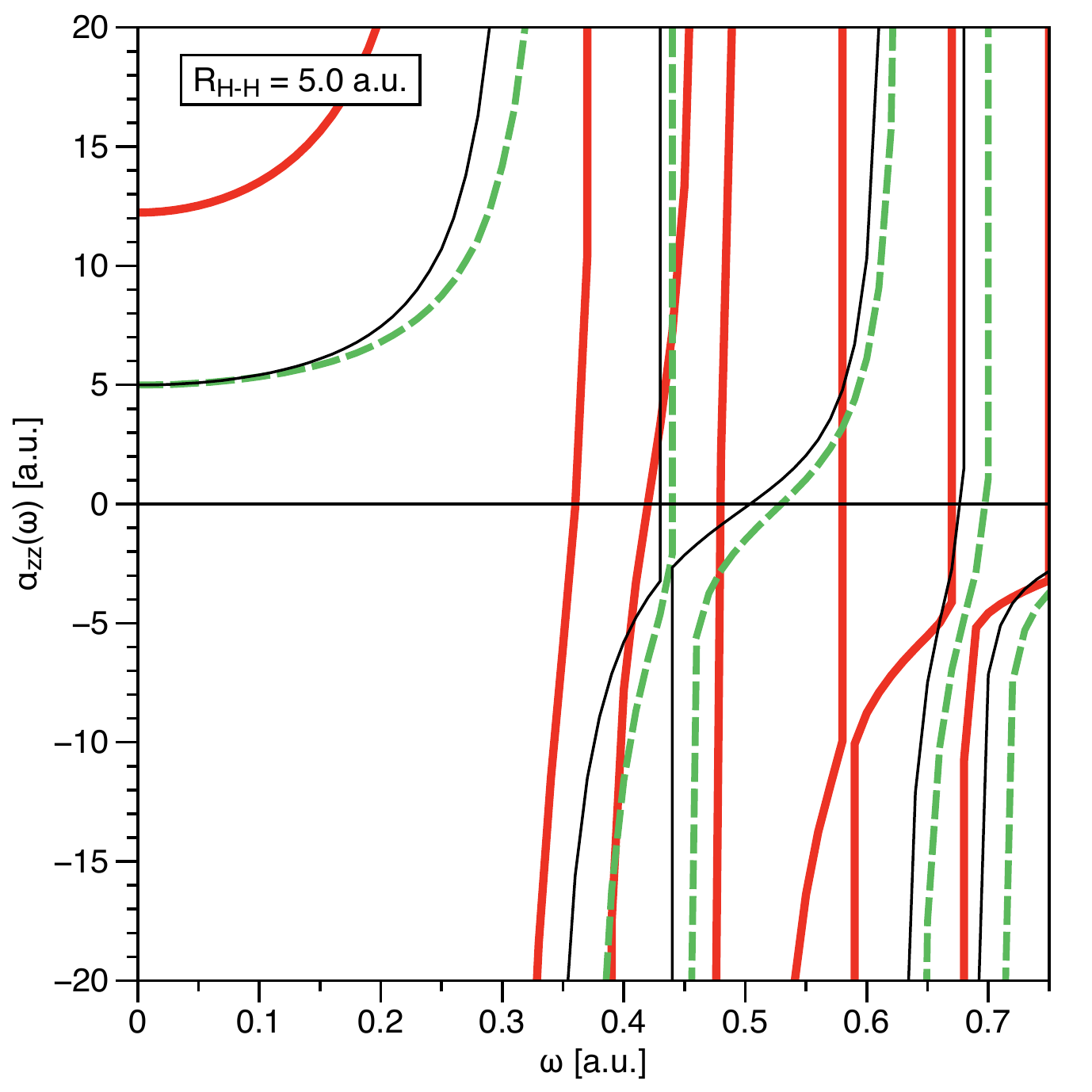}
  \caption{The $\alpha_{zz}(\omega)$ component of the polarizability tensor for H$_2$ at interatomic distances of 1.4 and 5.0 Bohr in an aug-cc-pVTZ basis. Exact (full CI): thick (red) lines; PILS (1 $\to$ all): dashed (green) lines; DMLS (1 $\to$ all): thin (black) lines.}
  \label{fig:H2_zz_1}
\end{figure}

The PILS functional, contrary to the DMLS functional, shows perfect agreement with the exact results, since it falls exactly on top of the full CI results. It can actually be shown analytically that the PINO response equations with PILS functional in fact constitute a reformulation of the full CI equations for the two-electron system, so this should indeed be the case. Further, although the DMLS functional shows a lot of spurious excitations (spikes) in the polarizability, it follows the exact polarizability rather closely on the $\omega$ intervals in between. Apparently, these spurious excitations carry little or no oscillator strength. Especially at equilibrium distance $R_{\text{H-H}} = R_e = 1.4$ Bohr, where double excitations are not important at this range of frequencies, the agreement of the DMLS curve with the exact and PILS curves is very good (apart from the poles).  However, for a stretched bond distance ($R_{\text{H-H}} = 5.0$ Bohr), double excitations become important and the agreement of the DMLS functional with the exact results at the intervals between the spikes  deteriorates (Fig.~\ref{fig:H2_zz_all}). Diagonal doubly excited nature of excited states (excitation from closed shell configuration $(\phi_i)^2$ to closed shell configuration $(\phi_a)^2$)  cannot be represented with the DMLS functional since, as a pure 1MFT functional, it suffers from the lack of response in the occupation numbers, which represents diagonal excited character~\cite{GiesbertzPernalGritsenko2009, GiesbertzGritsenkoBaerends2012b}. Such diagonal double excitation character ($(1\sigma_g)^2 \to (1\sigma_u)^2$) enters the low-lying excited states of H$_2$. 

\begin{figure}[t]
  \includegraphics[width=\columnwidth]{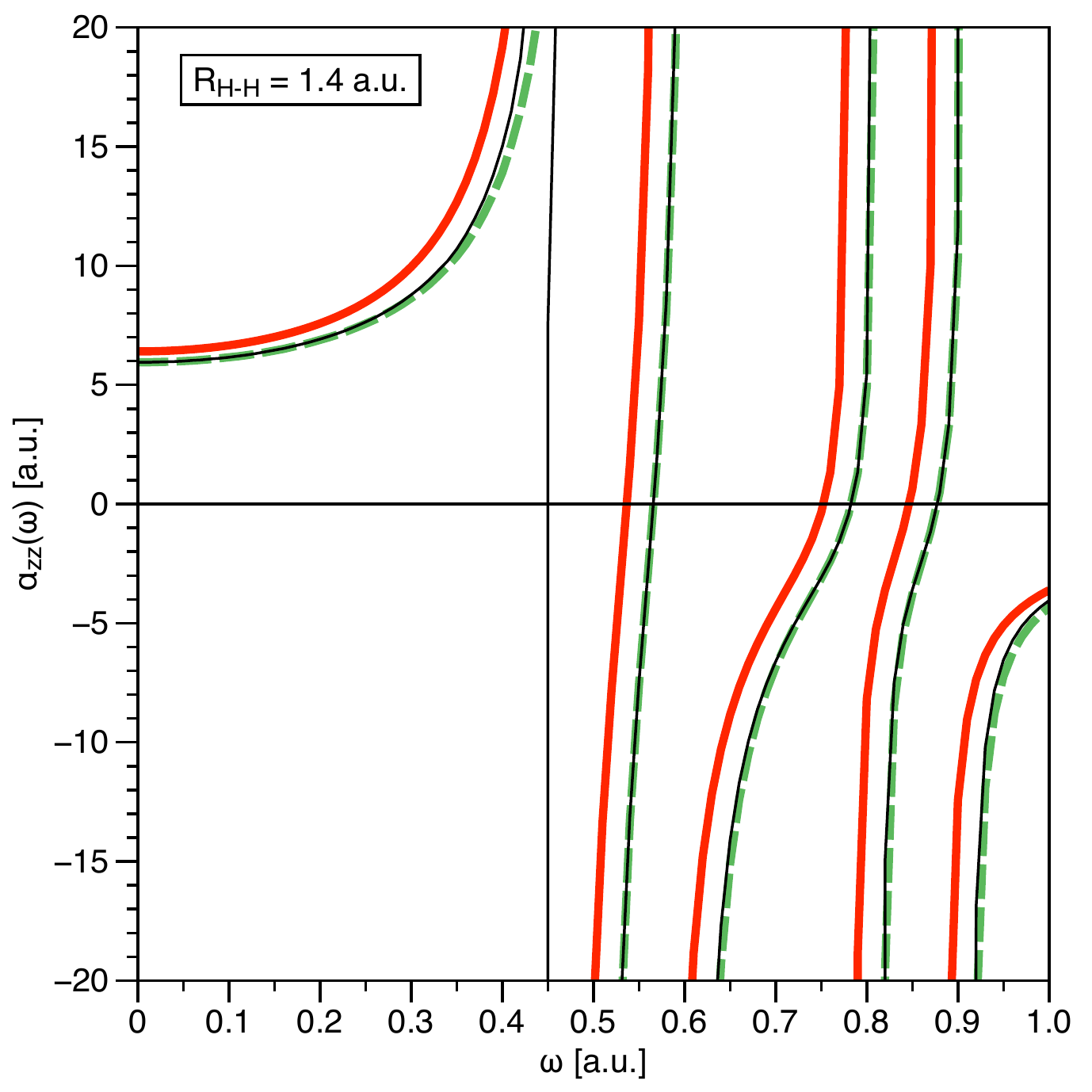}
  \includegraphics[width=\columnwidth]{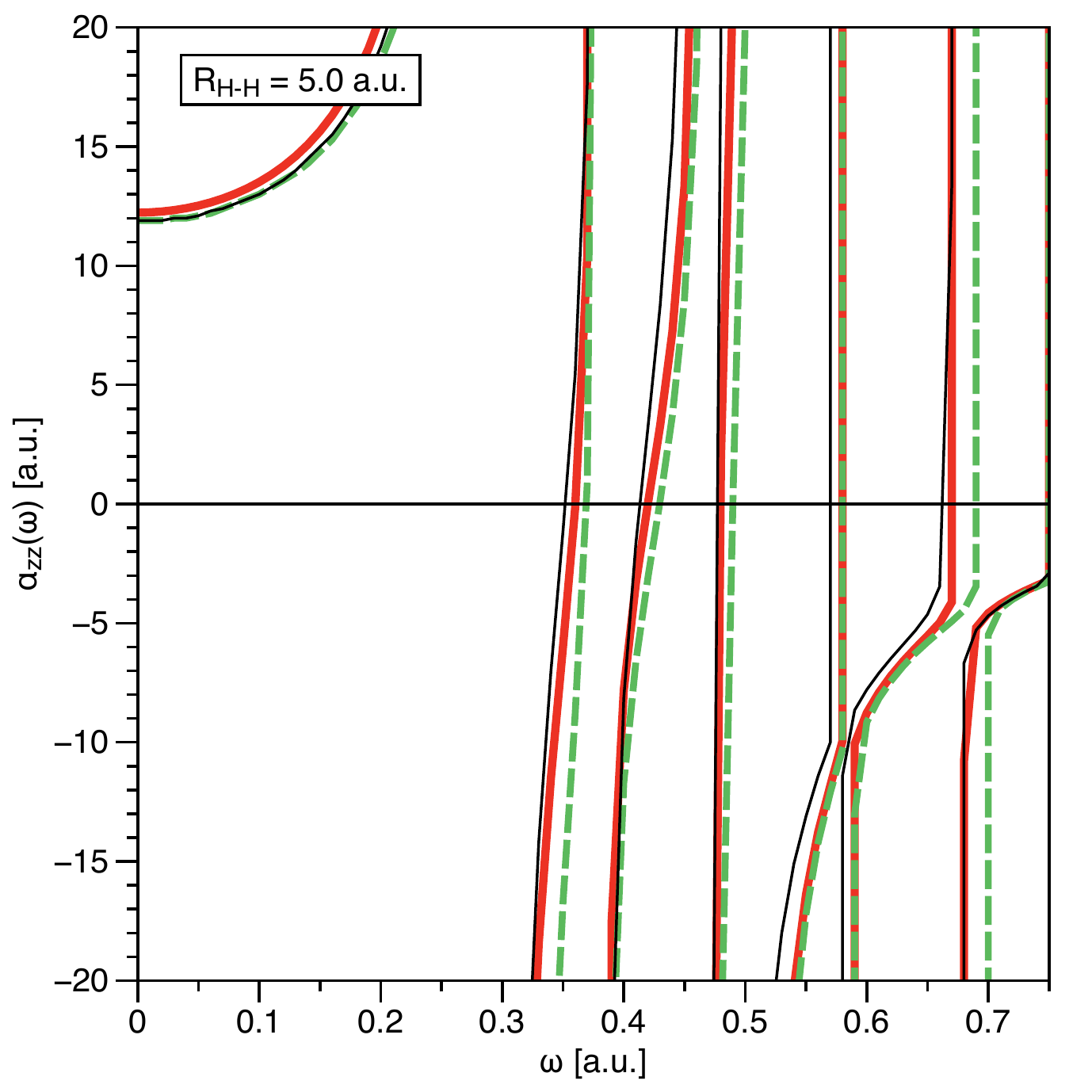}
  \caption{The $\alpha_{zz}(\omega)$ component of the polarizability tensor for H$_2$ at interatomic distances of 1.4 and 5.0 Bohr in an aug-cc-pVTZ basis. Exact (full CI): thick (red) lines; PILS (1, 2 $\to$ all): dashed (green) lines; DMLS (1, 2 $\to$ all): thin (black) lines.}
  \label{fig:H2_zz_1-2}
\end{figure}

One would expect that not all elements $\delta\gamma_{kl}$ are equally important. Since the $\delta \gamma_{kl}$ can be associated with $k \to l$ orbital transitions~\cite{GiesbertzGritsenkoBaerends2012b}, we would expect that notably virtual $\to$ virtual elements $\delta \gamma_{ab}$ where both $a$ and $b$ denote very weakly occupied NOs, would not be important for the polarizability (we denote weakly occupied NOs as ``virtuals"). Therefore, we also did a response calculation in which we only included transitions from the $1\sigma_g$ PINO to all virtuals (1 $\to$ all). This has been denoted the R0 variant in Ref.~\cite{GiesbertzGritsenkoBaerends2012b}, with roughly TDDFT size ($N_{\text{occ}}N_{\text{virt}} \times N_{\text{occ}}N_{\text{virt}}$ size of the response matrix). The diagonal elements $\delta \gamma_{pp}=\delta n_p$ are always included because of their importance for diagonal double excited nature of the states. The results for this calculation are shown in Fig.~\ref{fig:H2_zz_1}, where all the other parameters have been kept the same as in the previous calculation (Fig.~\ref{fig:H2_zz_all}). As expected, this limited calculation performs best at $R_e = 1.4$ Bohr, where the $1\sigma_g$ PINO has an occupancy of 0.98 and the $1\sigma_u$ PINO only an occupancy of 0.01. Especially transitions between PINOs with large occupancy difference should be important for low frequency fields\slash{}energies, since they give the most significant contributions to the lowest excitation energies~\cite{GiesbertzGritsenkoBaerends2012b, MeerGritsenkoGiesbertz2013}. Therefore, although only the transitions from $1\sigma_g$ PINO to the other ones have been included, we cover the most important ones, which is corroborated by the results in Fig.~\ref{fig:H2_zz_1} for 1.4 Bohr.

However, at a stretched bond distance of 5.0 Bohr, the $1\sigma_g$ occupation is only 0.63 and the $1\sigma_u$ occupation is increased to 0.37. Therefore, also ``transitions" from the $1\sigma_u$ PINO will be important for a correct calculation of response properties at low energies. This can be clearly seen from the second plot (5.0 Bohr) in Fig.~\ref{fig:H2_zz_1}, where the agreement for this R0 variant with the exact results is much poorer than at 1.4 Bohr.

Further note that if only the 1 $\to$ all excitations are included (R0) the DMLS results are rather close to the PILS results at 1.4 Bohr, and still reasonably close at 5.0 Bohr. They do not show any spurious spikes in the displayed interval at 1.4 Bohr, and only 1 at 5.0 Bohr. The spurious low lying excitations by the DMLS functional are off-diagonal double excitations which are represented by virtual-virtual transitions. Apparently, double excitations are made ``too easy'' by the DMLS functional, so they mix in at too low energies and produce many  spurious low-lying (double) excitations. By only allowing transitions from the $1\sigma_g$ PINO, we effectively remove all these bad virutal-virtual double excitations. However, they are required for the description of correlation, so the elimination of the spurious excitations comes at the cost that the exact polarizability as a function of the frequency is not followed so closely anymore: both PILS and DMLS differ considerably from the exact full CI (red) curves at 5.0 bohr. 

Since the $1\sigma_u$ PINO has such a large occupancy (0.37) at 5.0 Bohr, the removal of the transition out of the $1\sigma_u$ PINO in the R0 approximation is a severe limitation. Indeed, when we include also transitions from the 2\textsuperscript{nd} PINO (all elements $\delta \gamma_{2p}$ in the response of the 1RDM), the results at $R_{\text{H-H}} = 5.0$ Bohr improve significantly as shown in Fig.~\ref{fig:H2_zz_1-2}. The results for both the PILS and DMLS functional are now in very good agreement with the exact results at low frequencies [$\omega \lesssim 0.7$ a.u.]. Of course, also the results for $R_{\text{H-H}} = 1.4$ Bohr improve, but the improvement is not so spectacular as at $R_{\text{H-H}} = 5.0$ Bohr. The DMLS results already feature one spurious low excitation at 0.445 Hartree~\cite{PhD-Giesbertz2010}.

\section{Conclusion}
In this article we have studied the calculation of frequency dependent polarizabilities with the phase-including density matrix functional theory. The static limit ($\omega\to0$) of the frequency-dependent PINO response equations requires special attention to avoid unwarranted divergencies. We have shown that with careful treatment of the constant term in the perturbing potential, not only the static 1MFT response equations are recovered, but that also the perturbation in the chemical potential (the Lagrange multiplier for electron number conservation)  is treated correctly in this limit. The (spatially) constant time-dependent potential plays in the time-dependent PINO linear response equations a special role, which can be shown to be related to the perturbation in the chemical potential.

Further we have shown results for the $\alpha_{zz}(\omega)$ component of the polarizability tensor for H$_2$ at interatomic separations of 1.4 and 5.0 Bohr. The response calculation were performed with two different functionals: the PILS functional which explicitly depends on the PINO phases and the DMLS functional which is a proper 1RDM functional, so it does \emph{not} depend on the PINO phases. The PILS functional shows a perfect agreement with the exact results. It can be shown actually that the PILS functional can be regarded as a reformulation of the full CI equations for two-electron systems in the PINO basis, so these excellent results should be expected. The DMLS functional fails in the sense that it supports very many low lying spurious excitations. This is caused by the DMLS functional putting the double excitations at too low energies. At other $\omega$ values, it has also a good correspondence with the exact results at $R_{\text{H-H}} = 1.4$ Bohr.  At $R_{\text{H-H}} = 5.0$ Bohr, double excitations are more important, so the correspondence of the DMLS results withe the exact ones deteriorates.

Since the spurious low lying DMLS excitations are related to double excitations, the spectrum could be cleaned up by only including  excitations from the highest occupied PINO, the $1\sigma_g$, to the other PINOs. All spurious excitations could be removed this way and we obtained quite good results at $R_{\text{H-H}} = R_e = 1.4$ Bohr. However, this simplification of the calculations comes at a price: the results at 5.0 Bohr are no longer accurate, both in the DMLS calculations and also in the PILS calculations, with this approximation. At 5.0 Bohr, also the $1\sigma_u$ PINO has a large occupancy, so also excitations from this PINO should be important. Indeed, including also excitations from the $1\sigma_u$ PINO to all the others improved the results at 5.0 Bohr significantly for both the PILS and DMLS functional. Since the results were already quite accurate with the R0 variant at 1.4 Bohr, the inclusion of excitations from the $1\sigma_u$ PINO only slightly improved the results and the DMLS functional already shows one spurious low lying excitation.

These results show that the extension of 1MFT with functionals that include phase information is imperative when one wants to obtain reliable response properties. This will be important for developing  successful general $N$-electron PINO functionals.

\begin{acknowledgments}
This work has been supported by the Netherlands Foundation for Research (NWO): project 700-52-302 (KJHG and OVG) and a VENI grant 722.012.013 (KJHG) and by WCU (World Class University) program 
through the Korea Science and Engineering Foundation 
funded by the Ministry of Education, Science and Technology 
(Project No.\ R32-2008-000-10180-0) (KJHG, OVG and EJB).
\end{acknowledgments}

\appendix

\section{The sum-rule for $\mat{A}$}
\label{ap:AsumRule}
The sum-rule for the response matrix $\mat{A}$~\eqref{eq:Asumrule} can be derived in two steps. First we will show that $\mat{A}$ is hermitian in the sense that $A^{\vphantom{*}}_{kl,ba} = A^*_{ab,lk}$, after which the sum-rule is quite easy to prove. First we only consider the derivatives in the two-electron part of $\mat{A}$~\eqref{eq:Adef}, where we have to keep in mind that the response matrix is defined in a stationary basis PINO basis, so this transformation~\eqref{eq:dUdef} has to be taken into account
\begin{widetext}
\begin{align*}
K^{\phi}_{kl,ba} &\isDefinedAs
\sum_{rs}\integ{\vecx}\biggl(
\frac{\du U_{kr}\bigl(W^{\dagger}_{rs} - W^{\phantom{\dagger}}_{rs}\bigr)U^{\dagger}_{s l}}
{\du \pino_b(\vecx)}\pino_a(\vecx) - 
\frac{\du U_{kr}\bigl(W^{\dagger}_{rs} - W^{\phantom{\dagger}}_{rs}\bigr)U^{\dagger}_{s l}}{\du \pino^*_a(\vecx)}\pino^*_b(\vecx)\biggr) \\
&\hphantom{:}= \bigl(W^{\dagger}_{bl} - W^{\vphantom{\dagger}}_{bl}\bigr)\delta_{ka} -
\bigl(W^{\dagger}_{ka} - W^{\vphantom{\dagger}}_{ka}\bigr)\delta_{bl} \\*
&\hphantom{{}\isDefinedAs{}} {} +
\iinteg{\vecx}{\vecx'}
\biggl[\frac{\delta}{\delta\pino_{\!b}(\vecx')}\biggl(\pino^*_{\!k}(\vecx)\frac{\delta W}{\delta\pino^*_{\!l}(\vecx)} - 
\frac{\delta W}{\delta \pino_{\!k}(\vecx)}\pino_{\!l}(\vecx)\biggr)\pino_{\!a}(\vecx') -
\pino^*_{\!b}(\vecx')\frac{\delta}{\delta\pino^*_{\!a}(\vecx')}\biggl(\pino^*_{\!k}(\vecx)\frac{\delta W}{\pino^*_{\!l}(\vecx)} -
\frac{\delta W}{\delta\pino_{\!k}(\vecx)}\pino_{\!l}(\vecx)\biggr)\biggr] \\
&\hphantom{:}= K^{\phi:2}_{kl,ba} - W^{\vphantom{\dagger}}_{bl}\delta_{ka} - W^{\dagger}_{ka}\delta_{bl},
\end{align*}
where we introduced
\begin{align*}
K^{\phi:2}_{kl,ba} \isDefinedAs
\iinteg{\vecx}{\vecx'}\biggl(
&\pino^*_{\!k}(\vecx)\frac{\delta^2W}{\delta\pino^*_{\!l}(\vecx)\delta\pino_{\!b}(\vecx')}\pino_{\!a}(\vecx') -
\pino_{\!l}(\vecx)\frac{\delta^2W}{\delta\pino_{\!k}(\vecx)\delta\pino_{\!b}(\vecx')}\pino_{\!a}(\vecx') \\*
{} - {}&
\pino^*_{\!k}(\vecx)\frac{\delta^2W}{\delta\pino^*_{\!l}(\vecx)\delta\pino^*_{\!a}(\vecx')}\pino^*_{\!b}(\vecx') +
\pino_{\!l}(\vecx)\frac{\delta^2W}{\delta\pino_{\!k}(\vecx)\delta\pino^*_{\!a}(\vecx')}\pino^*_{\!b}(\vecx')\biggr).
\end{align*}
\end{widetext}
From the definition of $\mat{K}^{\phi:2}$ it is immediately clear that it is hermitian, $K^{\phi:2}_{kl,ba} = \bigl(K^{\phi:2}_{ab,lk}\bigr)^*$. The response matrix $\mat{A}$~\eqref{eq:Adef} can now be written as
\begin{align*}
A_{kl,ba} &= \bigl[n_bh_{ka} - \bigl(n_ah_{ka} + W^{\dagger}_{ka}\bigr)\bigr]\delta_{bl} \\*
&\eqspace {} +
\bigl[n_ah_{bl} - \bigl(n_bh_{bl} + W_{bl}\bigr)\bigr]\delta_{ka} + K^{\phi:2}_{kl,ba}.
\end{align*}
If we now work out the hermitian conjugate we have
\begin{align*}
A^*_{ab,lk} &= \bigl[n_lh^*_{ak} - \bigl(n_kh^*_{ak} + W_{ka}\bigr)\bigr]\delta_{lb} \\*
&\eqspace {} +
\bigl[n_kh^*_{lb} - \bigl(n_lh^*_{lb} + W^{\dagger}_{bl}\bigr)\bigr]\delta_{ak} + \bigl(K^{\phi:2}_{ab,lk}\bigr)^* \\
&= A_{kl,ba},
\end{align*}
where we used the stationarity condition
\begin{align*}
n_kh_{kl} + W_{kl} = n_lh_{kl} + W^{\dagger}_{kl}.
\end{align*}
This stationarity condition can be obtained from the EOM for the 1RDM~\eqref{eq:EOM1RDM} by requiring that the 1RDM is stationary, i.e.\ that all the time-derivatives vanish. This condition also follows from the first order stationarity conditions for the ground state state~\cite{Pernal2005, GiesbertzBaerends2010, GiesbertzGritsenkoBaerends2010b}.

To establish the sum-rule for $\mat{A}$~\eqref{eq:Asumrule}, we consider
\begin{align*}
\sum_kA_{kk,ba} &= \sum_k(n_b-n_a)h_{ba}(\delta_{bk} - \delta_{ka}) + \sum_kK^{\phi}_{kk,ba} \\
&= \sum_k\integ{\vecx}\biggl(
\frac{\du\bigl(W^{\dagger}_{kk} - W_{kk}\bigr)}{\du \pino_b(\vecx)}\pino_a(\vecx) \notag \\*
&\hphantom{\sum_k\integ{\vecx}\biggl(} {} -
\frac{\du\bigl(W^{\dagger}_{kk} - W_{kk}\bigr)}{\du \pino^*_a(\vecx)} \pino^*_b(\vecx)\biggr) = 0,
\end{align*}
where in the last step we used that $\Trace\{\mat{W}\} = 2W$, where $W$ was the two-electron part of the energy. The equality between the trace of $\mat{W}$ and twice the two-body part of the energy follows from~\cite{PernalCioslowski2005, PhD-Giesbertz2010}
\begin{align*}
W_{kl} = \integ{\vecx}\frac{\du W}{\du\pino_{\!k}(\vecx)}\pino_{\!l}(\vecx)
= \sum_{rst}\Gamma_{krst}w_{tsrl},
\end{align*}
where the two-body two-electron reduced density matrix (2RDM) is defined as
\begin{align*}
\Gamma(\vecx_1\vecx_2,\vecx_2'\vecx_1')
\isDefinedAs \brakket{\Psi}{\crea{\psi}(\vecx_1')\crea{\psi}(\vecx_2')
\anni{\psi}(\vecx_2)\anni{\psi}(\vecx_1)}{\Psi}.
\end{align*}
Now using the hermiticity of $\mat{A}$, the sum-rule~\eqref{eq:Asumrule} follows immediately.

\section{The sum-rules for $\mat{C}$ and $\overline{\mat{W}}$}
\label{ap:CandWsumRules}

We first prove the sum-rule for $\overline{\mat{W}}$ by starting from the following stationarity condition
\begin{align}\label{eq:occStatCond}
h_{kk} + \frac{\du W}{\du n_k} = \epsilon_k,
\end{align}
which can be obtained by requiring the PINOs to be stationary, i.e.\ $\pino_k(\vecx t) = \e^{-\im\epsilon_k t}\pino_k(\vecx)$. For fractionally occupied PINOs one can prove that $\epsilon_k = \epsilon$~\cite{GiesbertzBaerends2010}. Now we multiply this equation by $n_k$ and sum over $k$, which gives
\begin{align*}
\sum_rn_r\left(h_{rr} + \frac{\du W}{\du n_r}\right) = \sum_r n_r\epsilon_r.
\end{align*}
Differentiating this equations with respect to the occupation number $n_k$, we find
\begin{align*}
\left(h_{kk} + \frac{\du W}{\du n_k}\right) + \sum_r \frac{\du^2 W}{\du n_k\du n_r} n_r = \epsilon_k.
\end{align*}
Using the definition of $\overline{\mat{W}}$~\eqref{eq:WbarDef} and the stationarity condition~\eqref{eq:occStatCond} the sum-rule for $\overline{\mat{W}}$~\eqref{eq:CandWbarSumRules} immediately follows.

The sum-rule for $\overline{\mat{W}}$ can now be used to establish the sum-rule for $\mat{C}$ from the stationary response equations. Acting with $\mat{n}^T$ on the last set of stationary response equations~\eqref{eq:PINO_stat_dn} we find
\begin{align*}
0 &= \sum_rn_r \bigl(\delta v_{rr} - \delta\epsilon\bigr) - \sum_{rs} n_r\frac{\du^2 W}{\du n_r \du n_s}\delta n_r \\
&= \sum_r n_rC^T_{r,ab}\delta U^R_{ab}.
\end{align*}
Since this equation should hold for arbitrary $\delta\mat{U}^R$, we find the sum-rule for $\mat{C}$~\eqref{eq:CandWbarSumRules}.

\bibliography{bible}

\begin{thebibliography}{32}%
\makeatletter
\providecommand \@ifxundefined [1]{%
 \@ifx{#1\undefined}
}%
\providecommand \@ifnum [1]{%
 \ifnum #1\expandafter \@firstoftwo
 \else \expandafter \@secondoftwo
 \fi
}%
\providecommand \@ifx [1]{%
 \ifx #1\expandafter \@firstoftwo
 \else \expandafter \@secondoftwo
 \fi
}%
\providecommand \natexlab [1]{#1}%
\providecommand \enquote  [1]{``#1''}%
\providecommand \bibnamefont  [1]{#1}%
\providecommand \bibfnamefont [1]{#1}%
\providecommand \citenamefont [1]{#1}%
\providecommand \href@noop [0]{\@secondoftwo}%
\providecommand \href [0]{\begingroup \@sanitize@url \@href}%
\providecommand \@href[1]{\@@startlink{#1}\@@href}%
\providecommand \@@href[1]{\endgroup#1\@@endlink}%
\providecommand \@sanitize@url [0]{\catcode `\\12\catcode `\$12\catcode
  `\&12\catcode `\#12\catcode `\^12\catcode `\_12\catcode `\%12\relax}%
\providecommand \@@startlink[1]{}%
\providecommand \@@endlink[0]{}%
\providecommand \url  [0]{\begingroup\@sanitize@url \@url }%
\providecommand \@url [1]{\endgroup\@href {#1}{\urlprefix }}%
\providecommand \urlprefix  [0]{URL }%
\providecommand \Eprint [0]{\href }%
\providecommand \doibase [0]{http://dx.doi.org/}%
\providecommand \selectlanguage [0]{\@gobble}%
\providecommand \bibinfo  [0]{\@secondoftwo}%
\providecommand \bibfield  [0]{\@secondoftwo}%
\providecommand \translation [1]{[#1]}%
\providecommand \BibitemOpen [0]{}%
\providecommand \bibitemStop [0]{}%
\providecommand \bibitemNoStop [0]{.\EOS\space}%
\providecommand \EOS [0]{\spacefactor3000\relax}%
\providecommand \BibitemShut  [1]{\csname bibitem#1\endcsname}%
\let\auto@bib@innerbib\@empty
\bibitem [{\citenamefont {Giesbertz}\ \emph {et~al.}(2008)\citenamefont
  {Giesbertz}, \citenamefont {Baerends},\ and\ \citenamefont
  {Gritsenko}}]{GiesbertzBaerendsGritsenko2008}%
  \BibitemOpen
  \bibfield  {author} {\bibinfo {author} {\bibfnamefont {K.~J.~H.}\
  \bibnamefont {Giesbertz}}, \bibinfo {author} {\bibfnamefont {E.~J.}\
  \bibnamefont {Baerends}}, \ and\ \bibinfo {author} {\bibfnamefont {O.~V.}\
  \bibnamefont {Gritsenko}},\ }\href {\doibase 10.1103/PhysRevLett.101.033004}
  {\bibfield  {journal} {\bibinfo  {journal} {Phys. Rev. Lett.}\ }\textbf
  {\bibinfo {volume} {101}},\ \bibinfo {pages} {033004} (\bibinfo {year}
  {2008})}\BibitemShut {NoStop}%
\bibitem [{\citenamefont {Giesbertz}\ and\ \citenamefont
  {Baerends}(2008)}]{GiesbertzBaerends2008}%
  \BibitemOpen
  \bibfield  {author} {\bibinfo {author} {\bibfnamefont {K.~J.~H.}\
  \bibnamefont {Giesbertz}}\ and\ \bibinfo {author} {\bibfnamefont {E.~J.}\
  \bibnamefont {Baerends}},\ }\href {\doibase 10.1016/j.cplett.2008.07.018}
  {\bibfield  {journal} {\bibinfo  {journal} {Chem. Phys. Lett.}\ }\textbf
  {\bibinfo {volume} {461}},\ \bibinfo {pages} {338} (\bibinfo {year}
  {2008})}\BibitemShut {NoStop}%
\bibitem [{\citenamefont {Neugebauer}\ and\ \citenamefont
  {Baerends}(2004)}]{NeugebauerBaerends2004}%
  \BibitemOpen
  \bibfield  {author} {\bibinfo {author} {\bibfnamefont {J.}~\bibnamefont
  {Neugebauer}}\ and\ \bibinfo {author} {\bibfnamefont {E.~J.}\ \bibnamefont
  {Baerends}},\ }\href {\doibase 10.1063/1.1785775} {\bibfield  {journal}
  {\bibinfo  {journal} {J. Chem. Phys.}\ }\textbf {\bibinfo {volume} {121}},\
  \bibinfo {pages} {6155} (\bibinfo {year} {2004})}\BibitemShut {NoStop}%
\bibitem [{\citenamefont {Giesbertz}\ \emph {et~al.}(2009)\citenamefont
  {Giesbertz}, \citenamefont {Pernal}, \citenamefont {Gritsenko},\ and\
  \citenamefont {Baerends}}]{GiesbertzPernalGritsenko2009}%
  \BibitemOpen
  \bibfield  {author} {\bibinfo {author} {\bibfnamefont {K.~J.~H.}\
  \bibnamefont {Giesbertz}}, \bibinfo {author} {\bibfnamefont {K.}~\bibnamefont
  {Pernal}}, \bibinfo {author} {\bibfnamefont {O.~V.}\ \bibnamefont
  {Gritsenko}}, \ and\ \bibinfo {author} {\bibfnamefont {E.~J.}\ \bibnamefont
  {Baerends}},\ }\href {\doibase 10.1063/1.3079821} {\bibfield  {journal}
  {\bibinfo  {journal} {J. Chem. Phys.}\ }\textbf {\bibinfo {volume} {130}},\
  \bibinfo {pages} {114104} (\bibinfo {year} {2009})}\BibitemShut {NoStop}%
\bibitem [{\citenamefont {Pernal}\ \emph {et~al.}(2007)\citenamefont {Pernal},
  \citenamefont {Giesbertz}, \citenamefont {Gritsenko},\ and\ \citenamefont
  {Baerends}}]{PernalGiesbertzGritsenko2007}%
  \BibitemOpen
  \bibfield  {author} {\bibinfo {author} {\bibfnamefont {K.}~\bibnamefont
  {Pernal}}, \bibinfo {author} {\bibfnamefont {K.}~\bibnamefont {Giesbertz}},
  \bibinfo {author} {\bibfnamefont {O.}~\bibnamefont {Gritsenko}}, \ and\
  \bibinfo {author} {\bibfnamefont {E.~J.}\ \bibnamefont {Baerends}},\ }\href
  {\doibase 10.1063/1.2800016} {\bibfield  {journal} {\bibinfo  {journal} {J.
  Chem. Phys.}\ }\textbf {\bibinfo {volume} {127}},\ \bibinfo {pages} {214101}
  (\bibinfo {year} {2007})}\BibitemShut {NoStop}%
\bibitem [{\citenamefont {Appel}(2007)}]{Appel2007}%
  \BibitemOpen
  \bibfield  {author} {\bibinfo {author} {\bibfnamefont {H.}~\bibnamefont
  {Appel}},\ }\emph {\bibinfo {title} {Time-Dependent Quantum Many-Body
  Systems: Linear Response, Electronic Transport and Reduced Density
  Matrices}},\ \href@noop {} {Ph.D. thesis},\ \bibinfo  {school} {Freie
  Universit{\"a}t}, \bibinfo {address} {Berlin} (\bibinfo {year}
  {2007})\BibitemShut {NoStop}%
\bibitem [{\citenamefont {Giesbertz}\ \emph
  {et~al.}(2010{\natexlab{a}})\citenamefont {Giesbertz}, \citenamefont
  {Gritsenko},\ and\ \citenamefont
  {Baerends}}]{GiesbertzGritsenkoBaerends2010a}%
  \BibitemOpen
  \bibfield  {author} {\bibinfo {author} {\bibfnamefont {K.~J.~H.}\
  \bibnamefont {Giesbertz}}, \bibinfo {author} {\bibfnamefont {O.~V.}\
  \bibnamefont {Gritsenko}}, \ and\ \bibinfo {author} {\bibfnamefont {E.~J.}\
  \bibnamefont {Baerends}},\ }\href {\doibase 10.1103/PhysRevLett.105.013002}
  {\bibfield  {journal} {\bibinfo  {journal} {Phys. Rev. Lett.}\ }\textbf
  {\bibinfo {volume} {105}},\ \bibinfo {pages} {013002} (\bibinfo {year}
  {2010}{\natexlab{a}})}\BibitemShut {NoStop}%
\bibitem [{\citenamefont {Appel}\ and\ \citenamefont
  {Gross}(2010)}]{AppelGross2010}%
  \BibitemOpen
  \bibfield  {author} {\bibinfo {author} {\bibfnamefont {H.}~\bibnamefont
  {Appel}}\ and\ \bibinfo {author} {\bibfnamefont {E.~K.~U.}\ \bibnamefont
  {Gross}},\ }\href {\doibase 10.1103/PhysRevLett.90.043005} {\bibfield
  {journal} {\bibinfo  {journal} {Europhys. Lett.}\ }\textbf {\bibinfo {volume}
  {92}},\ \bibinfo {pages} {23001} (\bibinfo {year} {2010})}\BibitemShut
  {NoStop}%
\bibitem [{\citenamefont {Giesbertz}\ \emph
  {et~al.}(2010{\natexlab{b}})\citenamefont {Giesbertz}, \citenamefont
  {Gritsenko},\ and\ \citenamefont
  {Baerends}}]{GiesbertzGritsenkoBaerends2010b}%
  \BibitemOpen
  \bibfield  {author} {\bibinfo {author} {\bibfnamefont {K.~J.~H.}\
  \bibnamefont {Giesbertz}}, \bibinfo {author} {\bibfnamefont {O.~V.}\
  \bibnamefont {Gritsenko}}, \ and\ \bibinfo {author} {\bibfnamefont {E.~J.}\
  \bibnamefont {Baerends}},\ }\href {\doibase 10.1063/1.3499601} {\bibfield
  {journal} {\bibinfo  {journal} {J. Chem. Phys.}\ }\textbf {\bibinfo {volume}
  {133}},\ \bibinfo {pages} {174119} (\bibinfo {year}
  {2010}{\natexlab{b}})}\BibitemShut {NoStop}%
\bibitem [{\citenamefont {Cioslowski}\ and\ \citenamefont
  {Pernal}(2001)}]{CioslowskiPernal2001}%
  \BibitemOpen
  \bibfield  {author} {\bibinfo {author} {\bibfnamefont {J.}~\bibnamefont
  {Cioslowski}}\ and\ \bibinfo {author} {\bibfnamefont {K.}~\bibnamefont
  {Pernal}},\ }\href {\doibase 10.1063/1.1383292} {\bibfield  {journal}
  {\bibinfo  {journal} {J. Chem. Phys.}\ }\textbf {\bibinfo {volume} {115}},\
  \bibinfo {pages} {5784} (\bibinfo {year} {2001})}\BibitemShut {NoStop}%
\bibitem [{\citenamefont {Pernal}\ and\ \citenamefont
  {Baerends}(2006)}]{PernalBaerends2006}%
  \BibitemOpen
  \bibfield  {author} {\bibinfo {author} {\bibfnamefont {K.}~\bibnamefont
  {Pernal}}\ and\ \bibinfo {author} {\bibfnamefont {E.~J.}\ \bibnamefont
  {Baerends}},\ }\href {\doibase 10.1063/1.2137325} {\bibfield  {journal}
  {\bibinfo  {journal} {J. Chem. Phys.}\ }\textbf {\bibinfo {volume} {124}},\
  \bibinfo {pages} {014102} (\bibinfo {year} {2006})}\BibitemShut {NoStop}%
\bibitem [{\citenamefont {Pernal}\ and\ \citenamefont
  {Cioslowski}(2007)}]{PernalCioslowski2007}%
  \BibitemOpen
  \bibfield  {author} {\bibinfo {author} {\bibfnamefont {K.}~\bibnamefont
  {Pernal}}\ and\ \bibinfo {author} {\bibfnamefont {J.}~\bibnamefont
  {Cioslowski}},\ }\href {\doibase 10.1039/b704797e} {\bibfield  {journal}
  {\bibinfo  {journal} {Phys. Chem. Chem. Phys.}\ }\textbf {\bibinfo {volume}
  {9}},\ \bibinfo {pages} {5956} (\bibinfo {year} {2007})}\BibitemShut
  {NoStop}%
\bibitem [{\citenamefont {Requist}\ and\ \citenamefont
  {Pankratov}(2010)}]{RequistPankratov2010}%
  \BibitemOpen
  \bibfield  {author} {\bibinfo {author} {\bibfnamefont {R.}~\bibnamefont
  {Requist}}\ and\ \bibinfo {author} {\bibfnamefont {O.}~\bibnamefont
  {Pankratov}},\ }\href {\doibase 10.1103/PhysRevA.81.042519} {\bibfield
  {journal} {\bibinfo  {journal} {Phys. Rev. A}\ }\textbf {\bibinfo {volume}
  {81}},\ \bibinfo {pages} {042519} (\bibinfo {year} {2010})}\BibitemShut
  {NoStop}%
\bibitem [{\citenamefont {Giesbertz}\ \emph {et~al.}(2012)\citenamefont
  {Giesbertz}, \citenamefont {Gritsenko},\ and\ \citenamefont
  {Baerends}}]{GiesbertzGritsenkoBaerends2012b}%
  \BibitemOpen
  \bibfield  {author} {\bibinfo {author} {\bibfnamefont {K.~J.~H.}\
  \bibnamefont {Giesbertz}}, \bibinfo {author} {\bibfnamefont {O.~V.}\
  \bibnamefont {Gritsenko}}, \ and\ \bibinfo {author} {\bibfnamefont {E.~J.}\
  \bibnamefont {Baerends}},\ }\href {\doibase 10.1063/1.3687344} {\bibfield
  {journal} {\bibinfo  {journal} {J. Chem. Phys.}\ }\textbf {\bibinfo {volume}
  {136}},\ \bibinfo {pages} {094104} (\bibinfo {year} {2012})}\BibitemShut
  {NoStop}%
\bibitem [{\citenamefont {van Meer}\ \emph {et~al.}(2013)\citenamefont {van
  Meer}, \citenamefont {Gritsenko}, \citenamefont {Giesbertz},\ and\
  \citenamefont {Baerends}}]{MeerGritsenkoGiesbertz2013}%
  \BibitemOpen
  \bibfield  {author} {\bibinfo {author} {\bibfnamefont {R.}~\bibnamefont {van
  Meer}}, \bibinfo {author} {\bibfnamefont {O.~V.}\ \bibnamefont {Gritsenko}},
  \bibinfo {author} {\bibfnamefont {K.~J.~H.}\ \bibnamefont {Giesbertz}}, \
  and\ \bibinfo {author} {\bibfnamefont {E.~J.}\ \bibnamefont {Baerends}},\
  }\href {\doibase 10.1063/1.4793740} {\bibfield  {journal} {\bibinfo
  {journal} {J. Chem. Phys.}\ }\textbf {\bibinfo {volume} {138}},\ \bibinfo
  {pages} {094114} (\bibinfo {year} {2013})}\BibitemShut {NoStop}%
\bibitem [{\citenamefont {L{\"o}wdin}\ and\ \citenamefont
  {Shull}(1956)}]{LowdinShull1956}%
  \BibitemOpen
  \bibfield  {author} {\bibinfo {author} {\bibfnamefont {P.-O.}\ \bibnamefont
  {L{\"o}wdin}}\ and\ \bibinfo {author} {\bibfnamefont {H.}~\bibnamefont
  {Shull}},\ }\href {\doibase 10.1103/PhysRev.101.1730} {\bibfield  {journal}
  {\bibinfo  {journal} {Phys. Rev.}\ }\textbf {\bibinfo {volume} {101}},\
  \bibinfo {pages} {1730} (\bibinfo {year} {1956})}\BibitemShut {NoStop}%
\bibitem [{\citenamefont {Runge}\ and\ \citenamefont
  {Gross}(1984)}]{RungeGross1984}%
  \BibitemOpen
  \bibfield  {author} {\bibinfo {author} {\bibfnamefont {E.}~\bibnamefont
  {Runge}}\ and\ \bibinfo {author} {\bibfnamefont {E.~K.~U.}\ \bibnamefont
  {Gross}},\ }\href {\doibase 10.1103/PhysRevLett.52.997} {\bibfield  {journal}
  {\bibinfo  {journal} {Phys. Rev. Lett.}\ }\textbf {\bibinfo {volume} {52}},\
  \bibinfo {pages} {997} (\bibinfo {year} {1984})}\BibitemShut {NoStop}%
\bibitem [{\citenamefont {van Leeuwen}(1999)}]{Leeuwen1999}%
  \BibitemOpen
  \bibfield  {author} {\bibinfo {author} {\bibfnamefont {R.}~\bibnamefont {van
  Leeuwen}},\ }\href {\doibase 10.1103/PhysRevLett.82.3863} {\bibfield
  {journal} {\bibinfo  {journal} {Phys. Rev. Lett.}\ }\textbf {\bibinfo
  {volume} {82}},\ \bibinfo {pages} {3863} (\bibinfo {year}
  {1999})}\BibitemShut {NoStop}%
\bibitem [{\citenamefont {van Leeuwen}(2001)}]{Leeuwen2001}%
  \BibitemOpen
  \bibfield  {author} {\bibinfo {author} {\bibfnamefont {R.}~\bibnamefont {van
  Leeuwen}},\ }\href {\doibase 10.1142/S021797920100499X} {\bibfield  {journal}
  {\bibinfo  {journal} {Int. J. Mod. Phys. B}\ }\textbf {\bibinfo {volume}
  {15}},\ \bibinfo {pages} {1969} (\bibinfo {year} {2001})}\BibitemShut
  {NoStop}%
\bibitem [{\citenamefont {Ruggenthaler}\ and\ \citenamefont {van
  Leeuwen}(2011)}]{RuggenthalerLeeuwen2011}%
  \BibitemOpen
  \bibfield  {author} {\bibinfo {author} {\bibfnamefont {M.}~\bibnamefont
  {Ruggenthaler}}\ and\ \bibinfo {author} {\bibfnamefont {R.}~\bibnamefont {van
  Leeuwen}},\ }\href {\doibase 10.1209/0295-5075/95/13001} {\bibfield
  {journal} {\bibinfo  {journal} {Europhys. Lett.}\ }\textbf {\bibinfo {volume}
  {95}},\ \bibinfo {pages} {13001} (\bibinfo {year} {2011})}\BibitemShut
  {NoStop}%
\bibitem [{\citenamefont {L{\"o}wdin}(1955)}]{Lowdin1955a}%
  \BibitemOpen
  \bibfield  {author} {\bibinfo {author} {\bibfnamefont {P.-O.}\ \bibnamefont
  {L{\"o}wdin}},\ }\href {\doibase 10.1103/PhysRev.97.1474} {\bibfield
  {journal} {\bibinfo  {journal} {Phys. Rev.}\ }\textbf {\bibinfo {volume}
  {97}},\ \bibinfo {pages} {1474} (\bibinfo {year} {1955})}\BibitemShut
  {NoStop}%
\bibitem [{\citenamefont {Requist}\ and\ \citenamefont
  {Pankratov}(2011)}]{RequistPankratov2011}%
  \BibitemOpen
  \bibfield  {author} {\bibinfo {author} {\bibfnamefont {R.}~\bibnamefont
  {Requist}}\ and\ \bibinfo {author} {\bibfnamefont {O.}~\bibnamefont
  {Pankratov}},\ }\href {\doibase 10.1103/PhysRevA.83.052510} {\bibfield
  {journal} {\bibinfo  {journal} {Phys. Rev. A}\ }\textbf {\bibinfo {volume}
  {83}},\ \bibinfo {pages} {052510} (\bibinfo {year} {2011})}\BibitemShut
  {NoStop}%
\bibitem [{\citenamefont {Requist}(2012)}]{Requist2012}%
  \BibitemOpen
  \bibfield  {author} {\bibinfo {author} {\bibfnamefont {R.}~\bibnamefont
  {Requist}},\ }\href {\doibase 10.1103/PhysRevA.86.02211} {\bibfield
  {journal} {\bibinfo  {journal} {Phys. Rev. A}\ }\textbf {\bibinfo {volume}
  {86}},\ \bibinfo {pages} {022117} (\bibinfo {year} {2012})}\BibitemShut
  {NoStop}%
\bibitem [{\citenamefont {Vignale}(2008)}]{Vignale2008}%
  \BibitemOpen
  \bibfield  {author} {\bibinfo {author} {\bibfnamefont {G.}~\bibnamefont
  {Vignale}},\ }\href {\doibase 10.1103/PhysRevA.77.062511} {\bibfield
  {journal} {\bibinfo  {journal} {Phys. Rev. A}\ }\textbf {\bibinfo {volume}
  {77}},\ \bibinfo {pages} {062511} (\bibinfo {year} {2008})}\BibitemShut
  {NoStop}%
\bibitem [{\citenamefont {Giesbertz}(2010)}]{PhD-Giesbertz2010}%
  \BibitemOpen
  \bibfield  {author} {\bibinfo {author} {\bibfnamefont {K.~J.~H.}\
  \bibnamefont {Giesbertz}},\ }\emph {\bibinfo {title} {Time-Dependent One-Body
  Reduced Density Matrix Functional Theory; Adiabatic Approximations and
  Beyond}},\ \href@noop {} {Ph.D. thesis},\ \bibinfo  {school} {{V}rije
  {U}niversiteit}, \bibinfo {address} {De Boelelaan 1105, Amsterdam, The
  Netherlands} (\bibinfo {year} {2010})\BibitemShut {NoStop}%
\bibitem [{\citenamefont {M{\"u}ller}(1984)}]{Muller1984}%
  \BibitemOpen
  \bibfield  {author} {\bibinfo {author} {\bibfnamefont {A.~M.~K.}\
  \bibnamefont {M{\"u}ller}},\ }\href {\doibase 10.1016/0375-9601(84)91034-X}
  {\bibfield  {journal} {\bibinfo  {journal} {Phys. Lett. A}\ }\textbf
  {\bibinfo {volume} {105}},\ \bibinfo {pages} {446} (\bibinfo {year}
  {1984})}\BibitemShut {NoStop}%
\bibitem [{\citenamefont {Buijse}(1991)}]{PhD-Buijse1991}%
  \BibitemOpen
  \bibfield  {author} {\bibinfo {author} {\bibfnamefont {M.}~\bibnamefont
  {Buijse}},\ }\emph {\bibinfo {title} {Electron Correlation; Fermi and Coulomb
  holes dynamical and nondynamical correlation}},\ \href@noop {} {Ph.D.
  thesis},\ \bibinfo  {school} {{V}rije {U}niversiteit}, \bibinfo {address} {De
  Boelelaan 1105, Amsterdam, The Netherlands} (\bibinfo {year}
  {1991})\BibitemShut {NoStop}%
\bibitem [{\citenamefont {Buijse}\ and\ \citenamefont
  {Baerends}(2002)}]{BuijseBaerends2002}%
  \BibitemOpen
  \bibfield  {author} {\bibinfo {author} {\bibfnamefont {M.}~\bibnamefont
  {Buijse}}\ and\ \bibinfo {author} {\bibfnamefont {E.~J.}\ \bibnamefont
  {Baerends}},\ }\href {\doibase 10.1080/00268970110070243} {\bibfield
  {journal} {\bibinfo  {journal} {Mol. Phys.}\ }\textbf {\bibinfo {volume}
  {100}},\ \bibinfo {pages} {401} (\bibinfo {year} {2002})}\BibitemShut
  {NoStop}%
\bibitem [{\citenamefont {Dunning~Jr.}(1989)}]{cc-pVT/QZ_H_B-Ne_aug_H}%
  \BibitemOpen
  \bibfield  {author} {\bibinfo {author} {\bibfnamefont {T.~H.}\ \bibnamefont
  {Dunning~Jr.}},\ }\href {\doibase 10.1063/1.456153} {\bibfield  {journal}
  {\bibinfo  {journal} {J. Chem. Phys.}\ }\textbf {\bibinfo {volume} {90}},\
  \bibinfo {pages} {1007} (\bibinfo {year} {1989})}\BibitemShut {NoStop}%
\bibitem [{\citenamefont {Pernal}(2005)}]{Pernal2005}%
  \BibitemOpen
  \bibfield  {author} {\bibinfo {author} {\bibfnamefont {K.}~\bibnamefont
  {Pernal}},\ }\href {\doibase 10.1103/PhysRevLett.94.233002} {\bibfield
  {journal} {\bibinfo  {journal} {Phys. Rev. Lett.}\ }\textbf {\bibinfo
  {volume} {94}},\ \bibinfo {pages} {233002} (\bibinfo {year}
  {2005})}\BibitemShut {NoStop}%
\bibitem [{\citenamefont {Giesbertz}\ and\ \citenamefont
  {Baerends}(2010)}]{GiesbertzBaerends2010}%
  \BibitemOpen
  \bibfield  {author} {\bibinfo {author} {\bibfnamefont {K.~J.~H.}\
  \bibnamefont {Giesbertz}}\ and\ \bibinfo {author} {\bibfnamefont {E.~J.}\
  \bibnamefont {Baerends}},\ }\href {\doibase 10.1063/1.3426319} {\bibfield
  {journal} {\bibinfo  {journal} {J. Chem. Phys.}\ }\textbf {\bibinfo {volume}
  {132}},\ \bibinfo {pages} {194108} (\bibinfo {year} {2010})}\BibitemShut
  {NoStop}%
\bibitem [{\citenamefont {Pernal}\ and\ \citenamefont
  {Cioslowski}(2005)}]{PernalCioslowski2005}%
  \BibitemOpen
  \bibfield  {author} {\bibinfo {author} {\bibfnamefont {K.}~\bibnamefont
  {Pernal}}\ and\ \bibinfo {author} {\bibfnamefont {J.}~\bibnamefont
  {Cioslowski}},\ }\href {\doibase 10.1016/j.cplett.2005.06.103} {\bibfield
  {journal} {\bibinfo  {journal} {Chem. Phys. Lett.}\ }\textbf {\bibinfo
  {volume} {412}},\ \bibinfo {pages} {71} (\bibinfo {year} {2005})}\BibitemShut
  {NoStop}%
\end{thebibliography}%

\end{document}